\documentclass[12pt]{article} 

\usepackage{amsmath}
\usepackage{psfrag}
\usepackage[left=2cm,right=2cm,bottom=4cm, top=3cm]{geometry}

\usepackage{amsmath}
\usepackage{amsfonts}
\usepackage{mathtools}
\usepackage{comment}
\usepackage{color}
\usepackage[pdftex]{graphicx}
\usepackage{authblk}
\usepackage{bbold}

\title{Beyond the square root: Evidence for logarithmic dependence of market impact on size and participation rate}
\author[1]{Elia Zarinelli}
\author[2,3]{Michele Treccani}
\author[4,6]{J. Doyne Farmer}
\author[3,5,6]{Fabrizio Lillo}
\affil[1]{LIST S.p.A., via Carducci 20, Trieste, I-34127, Italy}
\affil[2]{Mediobanca S.p.A, Piazzetta E. Cuccia 1, 20121 Milano, Italy}
\affil[3]{QUANTLab, Via Pietrasantina 123, 56122 Pisa, Italy}
\affil[4]{Institute for New Economic Thinking at the Oxford Martin School and Mathematical Institute, Eagle House, Walton Well Rd., Oxford OX2 6ED, United Kingdom}
\affil[5]{Scuola Normale Superiore, Piazza dei Cavalieri 7, 56126 Pisa, Italy}
\affil[6]{Santa Fe Institute, 1399 Hyde Park Road, Santa Fe, NM 87501, USA}

\begin{document}

\maketitle

\begin{abstract}
We make an extensive empirical study of the market impact of large orders (metaorders) executed in the U.S. equity market between 2007 and 2009. We show that the square root market impact formula, which is widely used in the industry and supported by previous published research, provides a good fit only across about two orders of magnitude in order size.    A logarithmic functional form fits the data better, providing a good fit across almost five orders of magnitude.  We introduce the concept of an ``impact surface" to model the impact as a function of both the duration and the participation rate of the metaorder, finding again a logarithmic dependence. We show that during the execution the price trajectory deviates from the market impact, a clear indication of non-VWAP executions.  Surprisingly, we find that sometimes the price starts reverting well before the end of the execution. Finally we show that, although on average the impact relaxes to approximately $2/3$ of the peak impact, the precise asymptotic value of the price depends on the participation rate and on the duration of the metaorder. We present evidence that this might be due to a herding phenomenon among metaorders.

\end{abstract}

\newpage

\tableofcontents

\newpage
\section{Introduction}

The market impact\footnote{Also called price impact} of trades, i.e. the change in price conditioned on signed trade size, is a key property characterizing market liquidity and is important for understanding price dynamics \cite{bouchaud2008markets}.    As shown theoretically in the seminal work of Kyle \cite{kyle1985continuous}, the optimal strategy for an investor with private information about the future price of an asset is to trade incrementally through time.  This strategy allows earlier executions to be made at better prices and minimizes execution cost.
As done in recent papers, we will call the full orders {\it metaorders}  and the individual trades used to complete the execution {\it child orders}.  Here we study the market impact of metaorders and its dependence on other properties, such as participation rate and execution time. 

Kyle's original model \cite{kyle1985continuous} predicts that market impact should be a linear function of the metaorder size, but this requires a variety of idealized assumptions that may be violated in real markets.  Empirical studies have consistently shown that the market impact of a metaorder is a non-linear concave function of its size. The concave nature of market impact is robust, being observed for several heterogeneous datasets in terms of markets, epochs, and style of execution \cite{toth2011anomalous}.  Most earlier studies have concluded that the market impact of a metaorder is well described by  a ``square root law'' of market impact \cite{torre1997barra,almgren2005direct,engle2008measuring,toth2011anomalous, mastromatteo2014agent, brokmann2014,Iuga14}.  Defining market impact $\mathcal{I}$ as the expected average price return (or difference) between the the beginning and the end of a metaorder of size $Q$, the square-root law states that
\begin{equation}\label{eq_rms}
\mathcal{I}(Q) = \pm Y \sigma_D \left( \frac{Q}{V_D} \right)^{\delta}
\end{equation} 
where $\sigma_D$ is the daily volatility of the asset, $V_D$ is the daily traded volume, and the sign of the metaorder is positive (negative) for buy (sell) trades. The numerical constant $Y$ is of order unity and the exponent $\delta$ is in the range 0.4 to 0.7, but typically very close to $1/2$, i.e. to a square root.   Notice that the only conditioning variable is the total volume $Q$. This is surprising because it implies that the time taken to complete the metaorder and the participation rate are not individually important for explaining market impact -- the total order size $Q$ is all that matters.  

Most empirical studies of market impact make use of proprietary data of funds or brokerage firms, since the empirical analysis of metaorder's impact cannot be performed with public data. Therefore the vast majority of studies rely on a partial view of the market.   Exceptions are Refs. \cite{moro2009market,tothlillo10} where the whole market is considered and metaorders are reconstructed statistically from brokerage data.

In this paper we perform an extensive empirical investigation of the market impact of metaorders, relying on a dataset of several million metaorders executed in US equity markets on large, medium and small capitalisation stocks. The dataset is heterogeneous, containing metaorders traded by many financial institutions for different purposes, and it spans several years (in the present analysis we consider the period 2007- 2009).  The main strengths of our paper are the large number of metaorders and the heterogeneity of their origin. Market impact is very noisy and larger datasets can significantly help in reducing statistical uncertainty; our dataset has almost seven million metaorders, making it more than a factor of four larger than any previous study. Moreover the heterogeneity of institutions and brokers in this dataset guarantees that our results are not specific to a single execution strategy. For comparison in Table \ref{tab_number_0} we report the approximate number of metaorders investigated in previous literature. It is clear that our sample is more than an order of magnitude larger than the typical size investigated so far. Moreover, in contrast to other studies, the set of funds and brokers is large and heterogeneous. 

The main weakness of the dataset is that we have little knowledge and control on the conditions and characteristics of the execution. We do not know if the metaorders were executed for cash reasons or were informed trades (as in \cite{waelbroeck2013market}). Similarly, we do not know the execution algorithm used by the brokers (even if, as shown below, we can infer some information from the price dynamics during the metaorder execution). Finally, we do not know if trading size was conditioned on movement of the price during execution of the metaorder and if the daily metaorder was part of a longer execution over multiple days. All these effects can potentially bias the sample and have some role in the observed properties of the impact.

\begin{table}
\begin{center}
\begin{tabular}{|  l | r | r |}
\hline
Author & \# of metaorders & Institution \\
\hline
Almgren et al. \cite{almgren2005direct} & 700,000 & Citigroup\\
Engle et al. \cite{engle2008measuring} & 230,000 & Morgan Stanley\\
T\'oth et al.\cite{toth2011anomalous} &  500,000& CFM \\
Mastromatteo et al. \cite{mastromatteo2014agent} &  1,000,000& CFM \\
Brokmann et al. \cite{brokmann2014} &  1,600,000& CFM \\
Moro et al. \cite{moro2009market} &  150,000  & inferred\\
Bershova et al.  \cite{bershova2013non} &  300,000 & AllianceBernstein LP \\
Waelbroeck et al. \cite{waelbroeck2013market} &130,000& various \\
Bacry et al. \cite{Iuga14} & 400,000 & one broker\\
\hline  
\end{tabular}
\end{center}
\caption{The approximate number of metaorders considered in previous studies, together with the corresponding trading institution where the orders originated. }
\label{tab_number_0}
\end{table}

We do several things in this paper, studying the dependence of impact on the ratio of order size and volume as well as other conditioning variables,  the development of impact as a function of time, and the relaxation of price once the order is completed, as detailed below.

First, we test the limits of validity of the square root impact law by conditioning it on variables such as the market capitalization of the stock, the participation rate, and the duration of execution.  Because we are able to span more than five orders of magnitude of the ratio $Q/V_D$ in Equation \ref{eq_rms}, we are able to investigate deviations from the square root law more thoroughly than in previous studies.  Indeed we observe consistent deviations for large and small values of  $Q/V_D$, indicating that the power law relation of Eq. \ref{eq_rms} is only approximately valid.   Instead we find that a logarithmic function (which is more concave) fits the data significantly better.

Second, as suggested by a general class of market impact models, we study how market impact depends jointly on the duration $F$ and the participation rate $\eta$ of the metaorder.  Measuring time in units of traded volume, and letting $V_P$ be the volume exchanged by the whole market during the execution of the metaorder, the duration is the fractional volume $F = V_P/V_D$, and the participation rate  is the ratio of the order size $Q$ to the market volume while it is being executed, i.e. $\eta = Q/V_D$.  This implies that the conditioning variable $Q/V_D$ in Equation \ref{eq_rms} can be written as
\begin{equation}
\pi  \equiv \frac{Q}{V_D} = \frac{Q}{V_P} \frac{V_P}{V_D} = F \cdot \eta
\end{equation}
Thus the square root law of Eq. \ref{eq_rms} implicitly assumes that the impact depends only on the square root of the product of $F$ and $\eta$. We will show that this assumption is only approximate and that a more complex functional shape describes the data better. This functional form is described by an {\it impact surface} that takes into account the variation of market impact with both participation rate and execution time, or alternatively, any two of the three variables $\pi$, $\eta$ and $F$.   We show that the dependence of impact on these individual variables is better described by logarithms than power laws.

Third, we consider how the price changes {\it during} the execution of the metaorder.  Recent studies \cite{moro2009market,bershova2013non,waelbroeck2013market} find that impact is a concave function of time, i.e. for a given execution size, earlier transactions of the metaorder change the price more than later transactions.   By using a much larger dataset we confirm this observation, but we find that the pattern followed by the price during execution does not mirror the dependence of the metaorder on size. To say this more explicitly, consider two metaorders with the same participation rate, one with double the volume of the other. When the larger metaorder is halfway through its execution, will the impact at that point be equal to that of the smaller one that has just completed? The general answer is no:  The impact of the larger metaorder at that point in time will be larger than the impact of the smaller one. Interestingly, in some cases we find that the price starts reverting even before the end of the metaorder. We discuss some possible explanations for these findings.

Finally, the fourth question concerns the price dynamics after the conclusion of the metaorder. This topic (not covered in the original paper of Kyle \cite{kyle1985continuous}) has been receiving increasing attention recently \cite{moro2009market,farmer2013efficiency,bershova2013non,waelbroeck2013market,brokmann2014}.  Several studies indicate that once the metaorder is executed the market impact relaxes from its peak value and converges to a plateau \cite{moro2009market,bershova2013non,waelbroeck2013market}.  The reversion indicates that not all the impact is permanent.  Even stronger, a recent study suggests that, up to a proper deconvolution of the market impact with respect to the impact of subsequent metaorders and of the the price momentum, the impact relaxes to zero \cite{brokmann2014}.  

The measurement of permanent market impact is difficult for two reasons. First, the price after the end of the metaorder is very noisy, and a careful determination of the average price dynamics requires a large sample of metaorders. Second, if successive metaorders (whether by the same or different traders) are correlated in sign, it might be difficult to isolate the permanent impact of an individual metaorder \cite{brokmann2014}.

By making use of our large and heterogeneous sample, we perform careful measurements of the permanent impact of metaorders, considering different participation rates and durations.  For typical metaorder durations and participation rates we find that after the end of the metaorder the price decays to a value which {\it on average} is roughly $2/3$ of the peak impact, as suggested by \cite{farmer2013efficiency} and found empirically by \cite{bershova2013non}. However, we show that the measured price decay depends on the participation rate and duration of the metaorder. Based on empirical evidence, we postulate that this dependence can be in part explained by a herding phenomenon accounting for the fact that metaorders executed in the same time period tend to have similar sign (buy or sell). Thus correlation between the sign of nearby metaorders might be partly responsible for the level of the plateau reached by permanent impact. 

The paper is organized as follows. In Section \ref{sec:defdat} we present the definition of the variables and the averaging procedure. We also discuss the dataset and some descriptive statistics. Section \ref{sec_models} presents some models of the price dynamics during the execution of a metaorder, used later to understand the empirical findings. In Section \ref{sec:measurement} we present our empirical results and in Section \ref{fundamentalModels} we discuss the implications of our empirical results on fundamental models of market impact. Finally, in Section \ref{sec:conclusions} we draw some conclusions.


\section{Definitions and Data}\label{sec:defdat}

In this section we define the parameters we use to describe metaorder execution and the relative measures we consider to quantify market impact. In a second part we describe the database on which our analysis relies and we present some summary statistics of metaorders. 

\subsection{Definitions}


One of the well known facts of intraday financial data is the presence of very strong periodicities. In particular, the level of trading activity is known to vary substantially and consistently between different periods of the trading day, and this intra-day variation affects both the volume profile and the variance of prices. Therefore one minute at the opening is quite different, in terms of volume, from a minute in the middle of the day. In order to take into account the intraday patterns, in this paper we perform all our computations in \emph{volume time}. This consists in moving forward time according to the volume traded in the market. For a trading day, let $V(t)$ be the total volume traded by the market from the opening until (physical)  time $t$. We measure volume time via $v=v(t):= V(t)/V(t_c)$, where $t_c$ is the daily closing time and $V(t_c)$ is the volume traded in that day. The relationship between the physical time $t$ and the volume time $v$ is independent of the total daily volume. In particular, $v=0$ at market open and $v=1$ at market close. 
\\

We introduce three non-local parameters characterising the execution of a metaorder buying/selling ($\epsilon = \pm 1$) $Q$ shares in a physical time interval $[t_s,t_e]$. The \textbf{participation rate} $\eta$ is defined as the ratio between the volume $Q$ traded by the metaorder and the volume traded by the whole market during the execution interval
\begin{equation}
\eta := \frac{Q}{V(t_e)-V(t_s)}.
\end{equation}
The \textbf{duration} $F$ of a metaorder in volume time is defined by 
\begin{equation}
F := v(t_e)-v(t_s) = \frac{V(t_e)-V(t_s)}{V(t_c)}.
\end{equation}
The \textbf{daily fraction} $\pi$ is defined as the ratio between the volume $Q$ traded by the metaorder and the volume traded by the market in the whole day, i.e.  $\pi := {Q}/{V(t_c)}$. The metaorders we consider are executed within a single trading day, therefore these parameters are between $0$ and $1$. The three variables are clearly not independent, because it is $\pi = \eta \cdot F$.
\\

To quantify the market impact of the execution of a metaorder we define $s(v)$ as the logarithm of the price $S(v)$ at volume time $v$ rescaled by the daily volatility $\sigma_D$, i.e. $s(v) := {\log S(v)}/{\sigma_D}$.  Letting $\epsilon$ be the sign of the metaorder, and $\Omega$ be any set of information upon which the market impact is conditioned, the market impact at time $v$ of a metaorder that started at time $v_s<v$  is 

\begin{equation} 
\mathcal{I}(v | \Omega) :=  \mathbb{E}\left[ \left. \epsilon \left( s(v) -s(v_s) \right)  \right|  \Omega \right].
\label{eq_imp_v}
\end{equation}
We will consider conditioning sets $\Omega$ involving $\eta$, $F$ and $\pi$ as well as global information like the market capitalisation of the traded stock or the year when the metaorder is executed. With $\mathbb{E}\left[ \cdot \right|\Omega]$ we refer to the sample average over all metaorders belonging to the same set $\Omega$. 

We will consider three types of impact. The \textbf{immediate} market impact quantifies how  market impact builds up during the execution of the metaorder,  i.e. $v_s<v< v_e$. After the conclusion of the execution of the metaorder, $v>v_e$, the market impact relaxes toward the \textbf{permanent} market impact. The \textbf{temporary} market impact is measured at the moment $v=v_e$ when the metaorder is completed, i.e.
\begin{equation} \label{eq_imp_tmp}
\mathcal{I}_{tmp}(\Omega) := \mathbb{E}\left[ \left. \epsilon \left( s(v_e) -s(v_s) \right)  \right|  \Omega \right].
\end{equation}
The temporary market impact\footnote{This quantity is sometimes also called peak impact. Temporary impact should not be confused with the temporary component of impact, used for example in the Almgren-Chriss model \cite{almgren2001optimal}.} conditioned on the daily fraction $\pi$ defines the {\it market impact curve} $\mathcal{I}_{tmp}(\Omega=\{ \pi \} )$.  This is the quantity that has received the most attention in previous studies of market impact.  The temporary market impact conditioned on both the participation rate $\eta$ and the duration $F$ defines the {\it market impact surface}, $\mathcal{I}_{tmp}(\Omega=\{ \eta, F \} )$. The immediate market impact conditioned on both the participation rate $\eta$ and the duration $F$ defines the {\it market impact trajectory} $\mathcal{I}(v | \Omega=\{ \eta, F\})$, i.e. how the impact reaches the market impact surface $\mathcal{I}_{tmp}(\Omega=\{ \eta, F \} )$ during the execution of the metaorder.

\subsection{Metaorder execution data} \label{sec_meta}

Our analysis relies on the database made available by Ancerno, a leading transaction-cost analysis provider (\emph{www.ancerno.com})\footnote{ANcerno Ltd. (formerly the Abel Noser Corporation) is a widely recognised consulting firm that works with institutional investors to monitor their equity trading costs. Its clients include pension plan sponsors such as the California Public Employees' Retirement System (CalPERS), the Commonwealth of Virginia, and the YMCA retirement fund, as well as money managers such as MFS (Massachusetts Financial Services), Putman Investments, Lazard Asset Management, and Vanguard. Previous academic studies that use Ancerno data include \cite{puckett2008short,goldstein2009brokerage, chemmanur2009role, jame2010organizational, goldstein2011purchasing, puckett2011interim,busse2012buy}. In particular, the authors of \cite{puckett2008short} give evidence regarding the existence of weekly institutional herding, often resulting in intense buying and selling episodes which may affect the efficiency of security prices. The authors investigate the contemporaneous and subsequent abnormal returns of securities that institutional herds sell or buy. They bring evidence that stocks that herds buy outperform the stocks that herds sell prior to and during the week of portfolio formation. Then, intense sell herds are followed by return reversals while the contemporaneous returns associated with intense buy herds are permanent.}. 
The database contains data gathered by Ancerno on metaorder execution from the main investment funds and brokerage firms in the U.S.  For each metaorder we consider the stock symbol, the volume $Q$, the sign $\epsilon$, the starting time $t_s$ of the metaorder and the time $t_e$ when the metaorder is completed. Our analysis has been performed on a subset of the database, containing metaorders traded on the U.S. equity market from January 2007 to December 2009. Before filtering, this subset contains 28,386,564 metaorders.  This is more than an order of magnitude larger than any previous measurements of this kind (see table \ref{tab_number_0}). All metaorders are completed within one trading day. We introduce the following filters:
\begin{itemize}
\item \textbf{Filter 1}: we select the stocks which belong to the Russell3000 index. This filter is introduced in order to have the time series of the price for each analysed metaorder.  In this way we also discard metaorders executed on highly illiquid stocks.
\item \textbf{Filter 2}: we select metaorders ending before 4:01 PM. 
 
\item \textbf{Filter 3}: we select metaorders whose duration is longer than 2 minutes.

\item \textbf{Filter 4}: we select metaorders whose participation rate $\eta$ is smaller than $0.3$

\end{itemize}

\begin{table}
\begin{center}
\begin{tabular}{|  l | r | r | r | r | r |}
\hline
year & raw & Filter 1 & Filter 2 & Filter 3 & Filter 4 \\
  \hline                       
  2007 &  9,216,333 & 6,904,656 &  3,082,767 &  2,130,045 & 1,976,382  \\
  2008 &  9,955,238 & 8,074,103  & 4,035,043 &  2,731,572 & 2,563,674 \\
  2009 &  9,214,993 & 7,622,703 &  3,954,355 &  2,552,092 & 2,404,827 \\
  \hline 
  tot & 28,386,564 & 22,601,462 & 11,072,165 & 7,413,709 & 6,944,883 \\
  \hline  
\end{tabular}
\end{center}
\caption{Number of metaorders surviving each filter introduced in the analysis.}
\label{tab_number}
\end{table}

\begin{table}
\begin{center}
\begin{tabular}{| r | r | r | r | r | r | r | r | r | r |}
\hline
 JPM   & XOM  & MSFT  &  GE  &  PG &  BAC & CSCO  & AAPL &    T  &  GS  \\ 
    \hline  
37,179 & 36,676 & 36,112 & 35,490 & 34,216 & 34,163 & 33,750 & 32,007 & 31,652 & 30,921 \\
  \hline
    \hline
 QCOM   &  HPQ   &  WMT   &   VZ    & MRK     &  C  &   PFE   & GOOG   &  SLB   &  JNJ \\    
 \hline
30,765  & 29,915  & 29,898 & 27,975  & 27,106  & 26,767  & 26,392  & 26,202  & 25,821  & 25,602 \\
  \hline                       

\end{tabular}
\end{center}
\caption{Ticker symbol of the most traded stocks and corresponding number of metaorders.}
\label{tab_top}
\end{table}

The number of metaorders surviving each filter is reported in table \ref{tab_number}. In table \ref{tab_top} we present the stock symbols with the largest number of metaorders in the dataset and their number. It is interesting to note that for the top 20 stocks the metaorders recorded in the Ancerno database are responsible of around 5\% of the daily volume.
It is evident that Filter 2 cuts a significant fraction of metaorders. All these metaorders last exactly 410 minutes, starting at 9:30 AM and ending at 4:20 PM. A detailed investigation of these orders strongly suggests that the initial and final time of these orders are not reliable, and we suspect that for these orders the times communicated to Ancerno are not accurate. For example, these orders have systematically lower participation rate than the other orders, suggesting an effective shorter time span of execution. In order to avoid introducing data that might be spurious, we drop these metaorders, at the cost of significantly reducing our sample. 
\\

In conclusion, for each metaorder in the dataset we recover the relative daily fraction $\pi$, the participation rate $\eta$, and the duration $F$. By exploiting the price data, we also recover the time series of the price $s(v)$ during and after the execution of the metaorder.

\subsection{Market price data}

In order to augment the information in the metaorder data described in the previous subsection we augment it with market data. The latter are historical data provided by Kibot (\emph{www.kibot.com}),consisting of one-minute time series giving the  Date, Time, Open, High, Low, Close, Volume of $3,500$ stocks in the Russell3000 index. We consider as a proxy of the daily volatility $\sigma_D= (S_h-S_l)/S_o$, where $S_{h,l,o}$ are the high-low-open price of the day.  Given the time interval $[t_s,t_e]$ and the volume $Q$ of a metaorder, this dataset makes it possible to compute its participation rate $\eta$, daily rate $\pi$, and duration $F$. We also use this database to measure the price dynamics during and after the execution of the metaorder.

\subsection{Metaorder statistics}

\begin{figure}[t] 
  \centering
 	\includegraphics[width=1\textwidth]{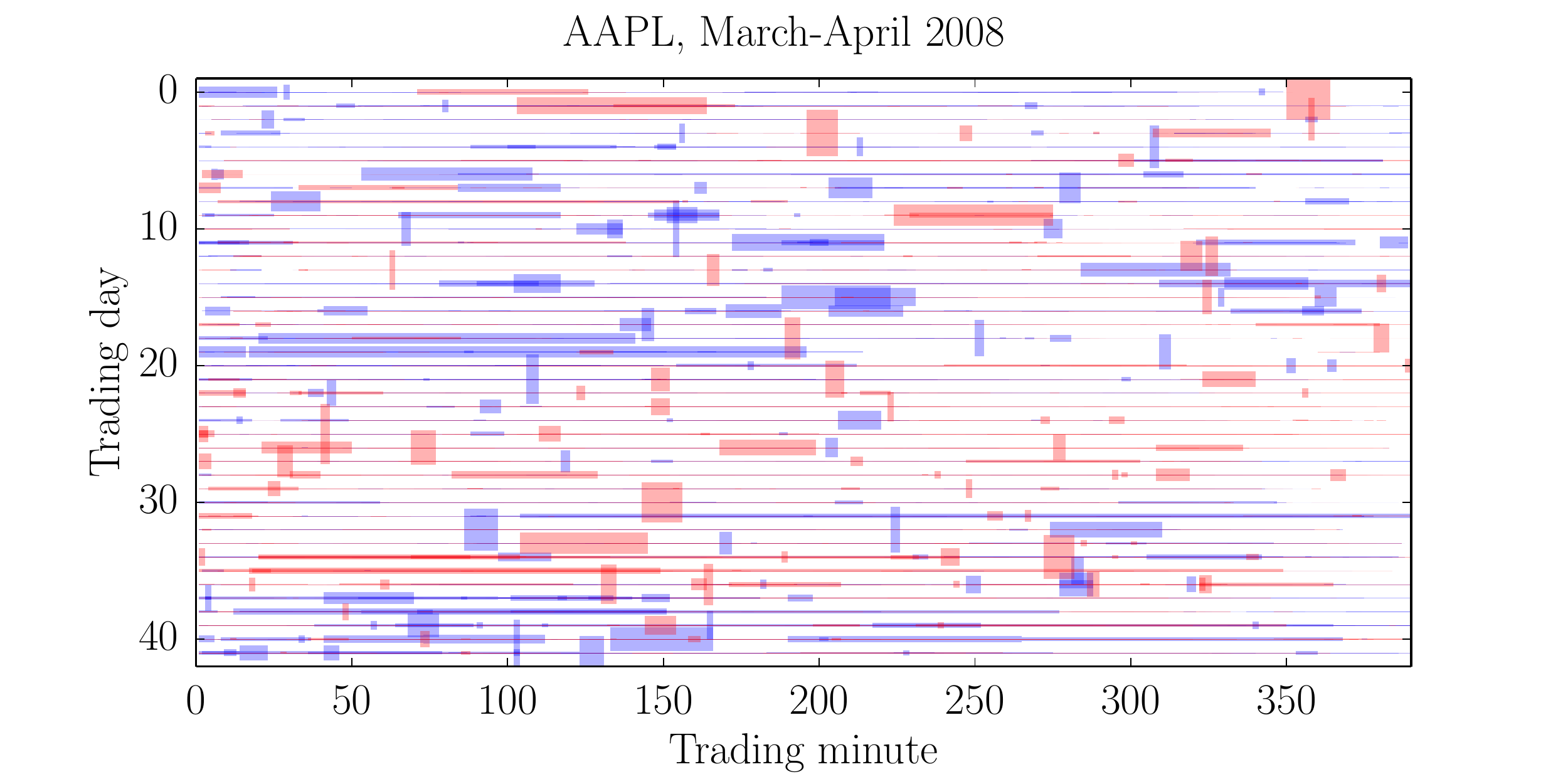}
  \caption{Time series of metaorders active on the market for AAPL in the period March-April 2008. Buy (Sell) metaorders are depicted in blue (red). The thickness of the line is proportional to the metaorder participation rate. More metaorders in the same instant of time give rise to darker colours. Each horizontal line is a trading day. We observe very few blanks, meaning that there is almost always an active metaorder from our database, which is of course only a subset of the number of orders that are active in the market.}
  \label{fig_aapl}
\end{figure}

We now present some descriptive statistics. In Figure \ref{fig_aapl} we show the time series of the metaorders for Apple (AAPL) in the period March-April 2008.  There is a significant number of metaorders active every day.  In most trading days there is a ``mood", i.e. on any given day most metaorders have the same sign, indicating a possible herding effect. Later we will quantify this metaorder overlap and we will discuss its possible role on the shape of market impact.

\begin{figure}[t] 
  \centering
 	\includegraphics[width=0.4\textwidth]{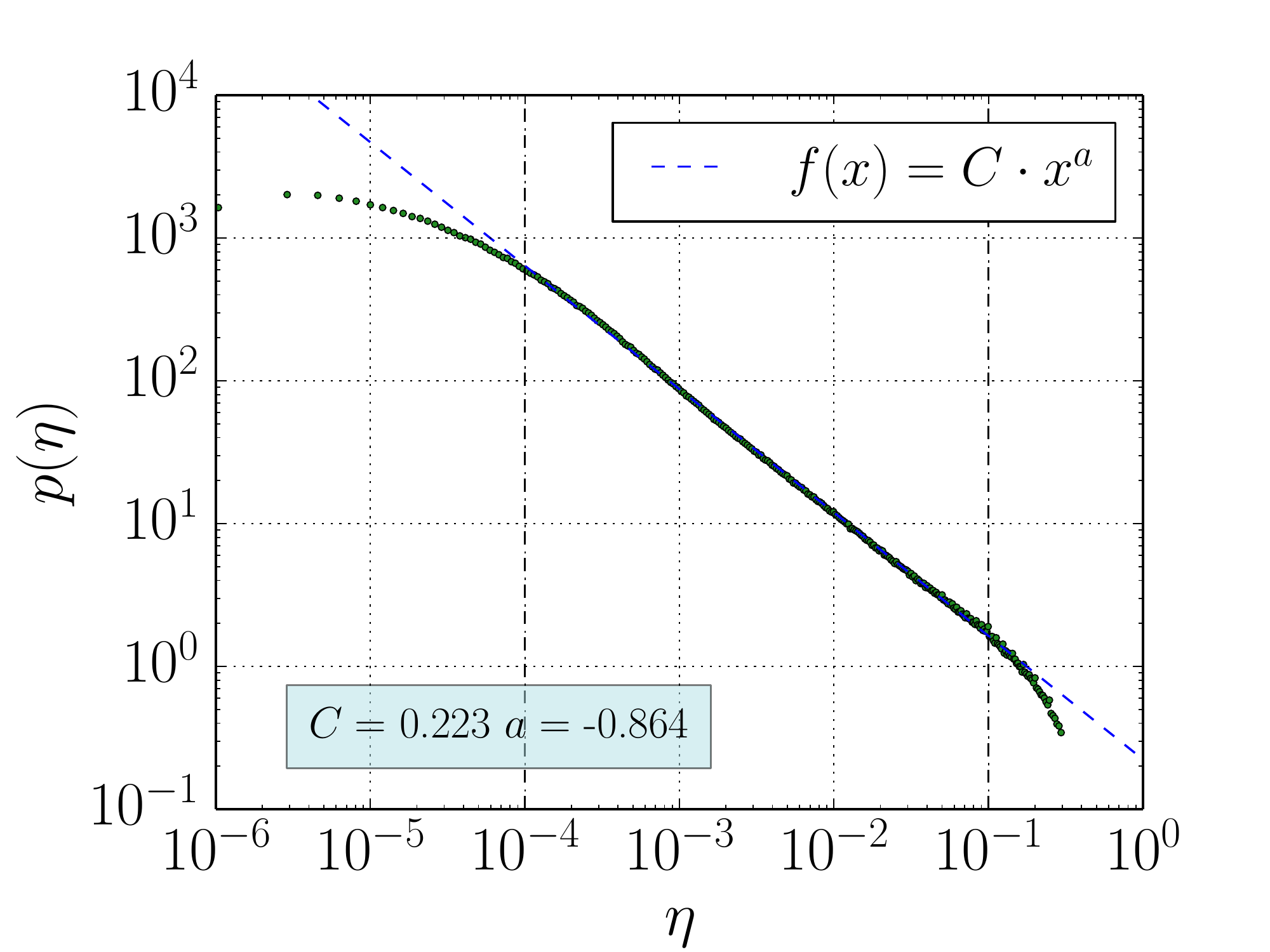}
	\includegraphics[width=0.4\textwidth]{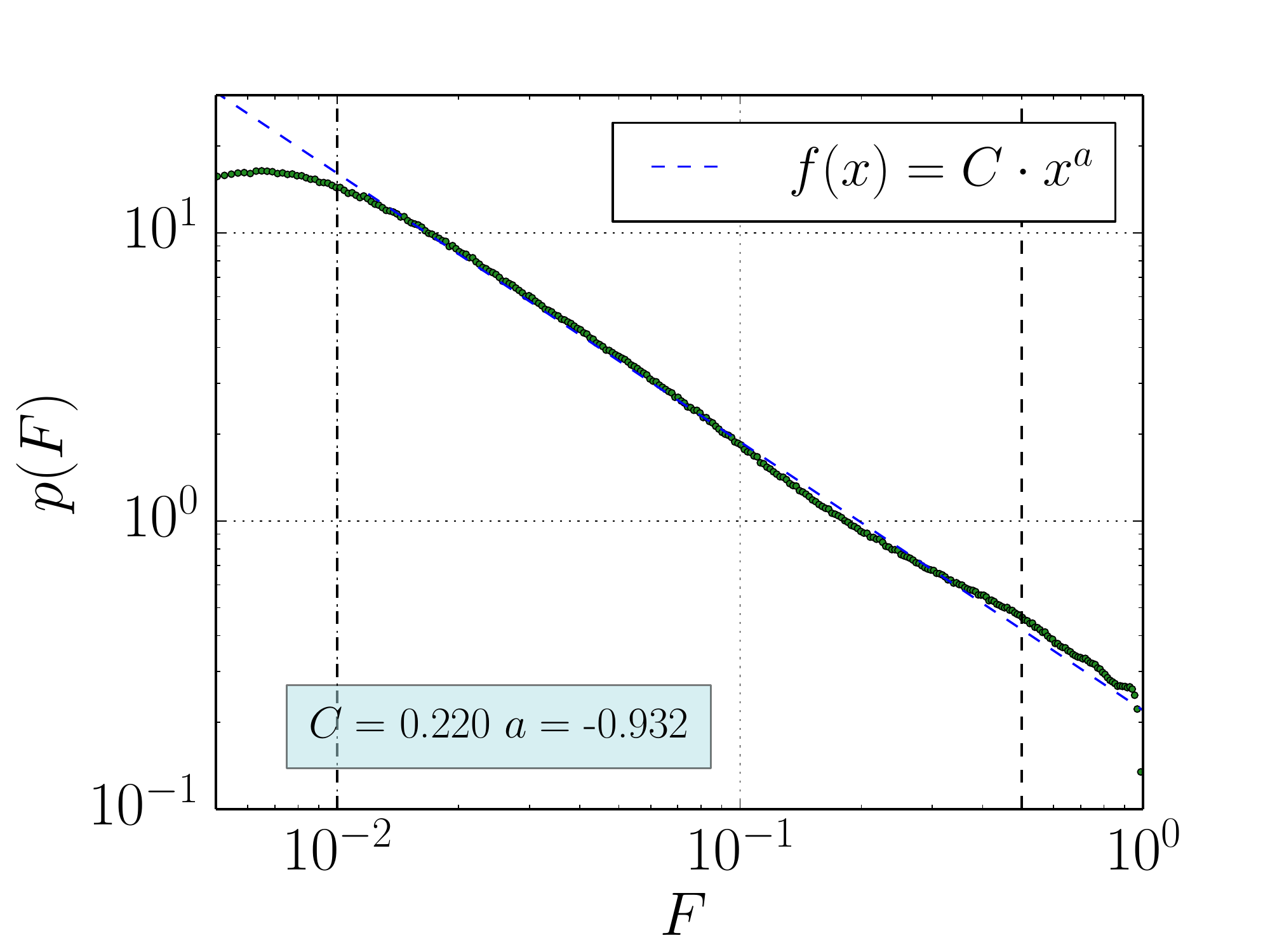}\\
	\includegraphics[width=0.4\textwidth]{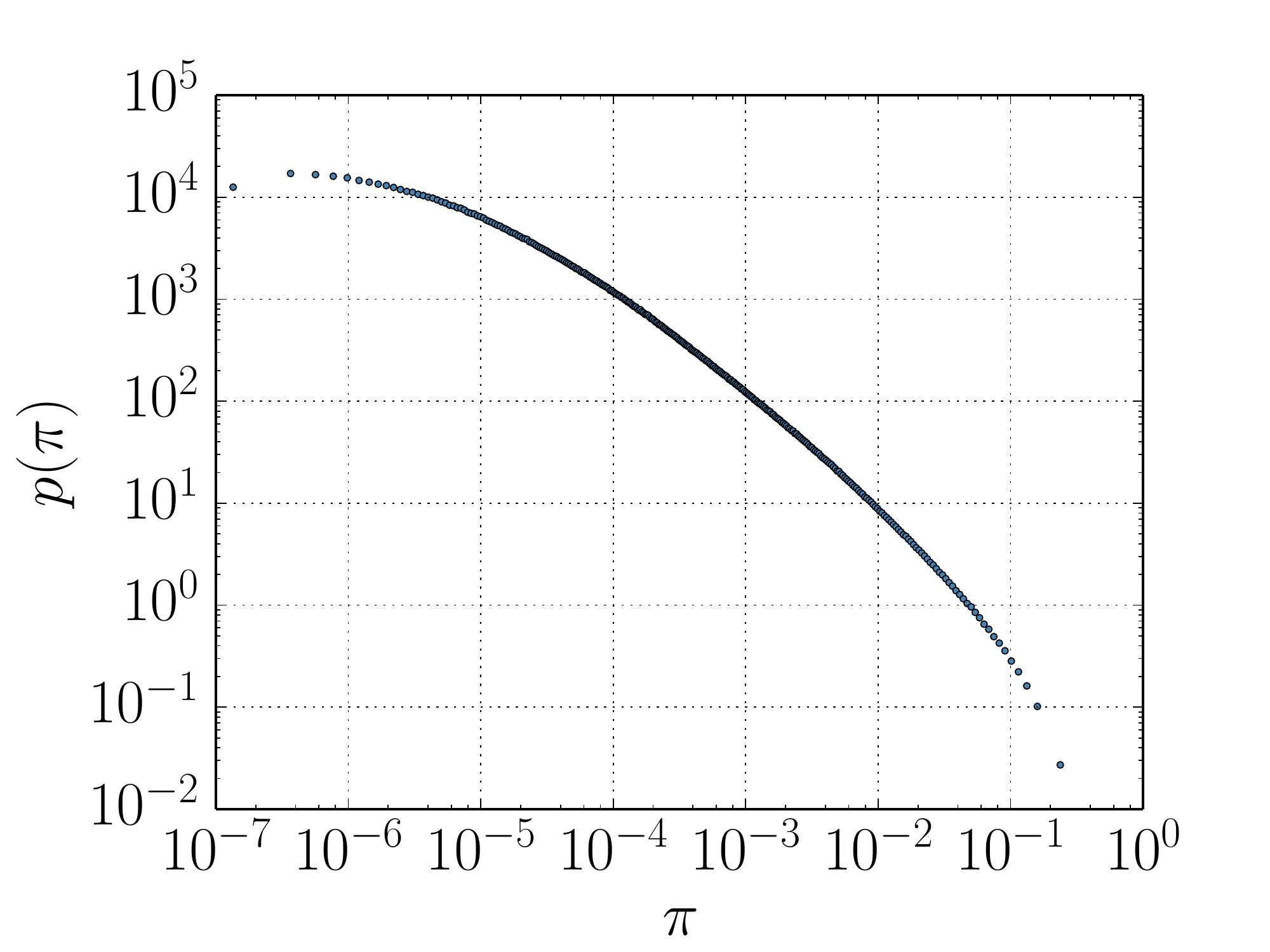}
	\includegraphics[width=0.4\textwidth]{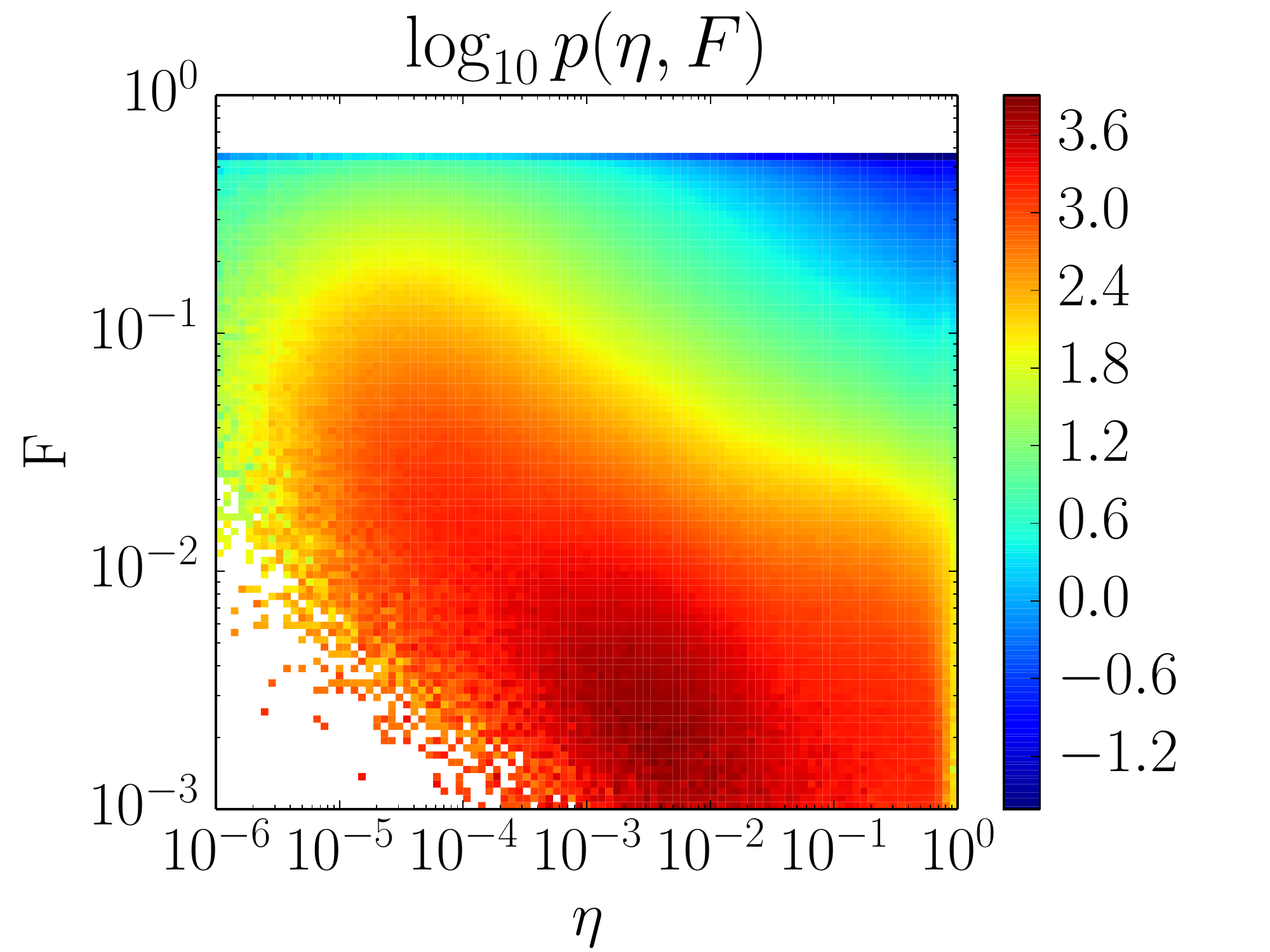}
  \caption{Estimation of the probability density function of the participation rate $\eta$ (top left), duration $F$ (top right), and daily fraction $\pi$ (bottom left). All these panels are in log-log scale and the first two shows also the best fit with a power law function in the region bounded by the two vertical dashed lines. The bottom right panel shows the logarithm of the estimated joint probability density function $p(\eta,F)$ in double logarithmic scale of the duration $F$ and the participation rate $\eta$. 
 }
  \label{fig_stat}
\end{figure}

We then investigate the distributional properties of the parameters characterising metaorders, namely the participation rate $\eta$, the duration $F$, and the daily fraction $\pi$. The statistics are performed by aggregating the  $6,944,883$ filtered metaorders.   

We find that the participation rate $\eta$ and the duration $F$ are both well approximated by a truncated power-law distribution over several orders of magnitude. The estimated probability density function of the participation rate $\eta$ is shown in log-log scale in the top left panel of Figure \ref{fig_stat}. A power law fit in the region $10^{-4}\le \eta  \le 0.1$, i.e. over three orders of magnitude gives a best fit exponent $a=-0.864 \pm 0.001$.   The top right panel of Figure \ref{fig_stat} shows the estimated probability density function of the duration $F$ of a metaorder.   A power law fit in the intermediate region bounded by the two vertical dashed lines ($0.01\le F \le 0.5$) gives a power-law exponent $a=-0.932 \pm 0.003$.  Thus in both cases the power law is very heavy tailed, meaning that there is substantial variability in both the partition rate and duration of the orders over a large range.  Note that in both cases the variability is intrinsically bounded (and therefore the power law is automatically truncated) by the fact that by definition $\eta \le 1$ and $F \le 1$.   In addition, for $p(F)$, there is a small bump on the right extreme of the distribution corresponding to all-day metaorders.  The deviation from a power law for small $F$ is forced automatically by our filter retaining only orders lasting at least $2$ minutes, which in volume time corresponds on average to $2/390\simeq 0.005$. 

The bottom left panel shows the probability density function of the daily fraction $\pi$. In this case the distribution is less fat tailed, and in particular it is clearly not a power law.  This is potentially an important result, as the predictions of some theories for market impact depend on this, and have generally assumed power law behavior \cite{gabaix06,farmer2013efficiency}.  

Since two of the three variables characterizing a metaorder are sufficient to derive the third one ($\pi = \eta \cdot F$), it is important to study the correlation between them, especially in light of the multivariate regression we perform below. The bottom right panel of Figure \ref{fig_stat} shows the logarithm of the estimated joint probability density function $p(\eta,F)$ in double logarithmic scale as a function of the duration $F$ and the participation rate $\eta$.
The linear correlation between the two variables is very low ($-0.022 $). The main contribution coming from the extreme regions, i.e. $\eta$ very large, implies $F$ very small and vice versa.
This means, as expected, that very aggressive metaorders are typically short and long metaorders more often have a small participation rate.


\section{Heuristic models of market impact} \label{sec_models}

Before presenting our empirical results on market impact, we consider some simple heuristic models of price dynamics.  By ``heuristic" we mean that these are reduced form models that are chosen because they are intuitively reasonable and they are useful, e.g. for computing optimal trade execution strategies.  We distinguish these from more fundamental models that try to explain the form of the market impact function from first principles.  These models provide a useful framework to investigate and to interpret our measures of market impact, presented in the next Section.  In particular it will provide a context to interpret some of the non-intuitive aspects of the relationship between immediate impact and temporary impact.  In Section \ref{fundamentalModels} we return to discuss some of the implications of our empirical work for fundamental models.

\subsection{The Almgren-Chriss model} \label{sec_ac}

\begin{figure}[t] 
  \centering
 	\includegraphics[width=0.53\textwidth]{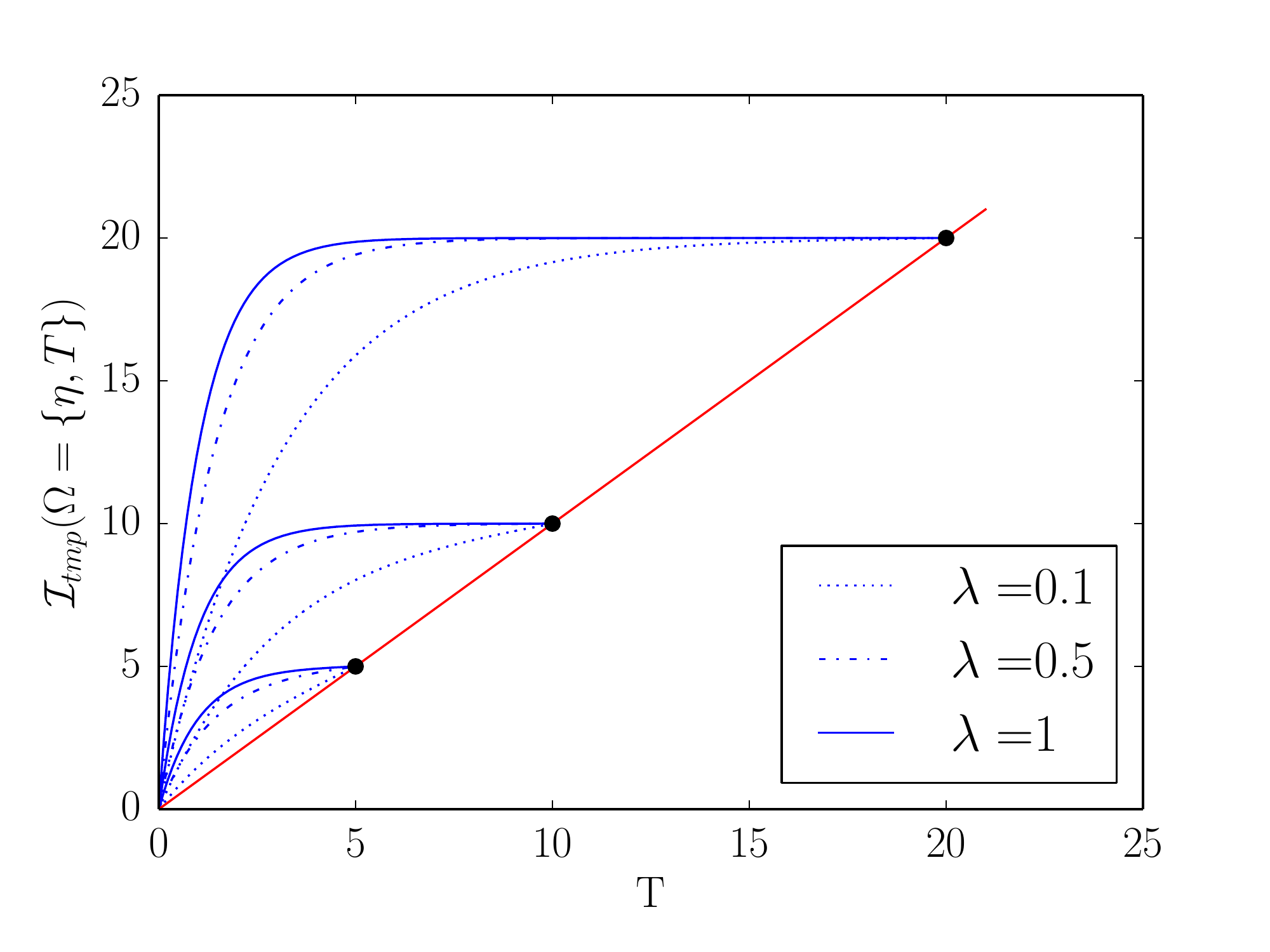}
  \caption{ Temporary market impact $\mathcal{I}_{tmp}(\Omega=\{ \eta, T \})$ as a function of the metaorder duration $T$ evaluated in the framework of the simplified version of the Almgren-Chriss model with $a=1$, $\sigma =1$ and $\eta =1$ (red line). We show also the immediate market impact trajectory $\mathcal{I}(t|\Omega=\{ \eta, T \})$  for several values of the risk-aversion parameter $\lambda = \{ 0.1,0.5,1\}$ and several metaorder durations $T=\{ 5,10,20 \}$ (blue lines).}
  \label{fig_ac_0}
\end{figure}

We consider first a simplified version of the Almgen-Chriss model \cite{almgren2001optimal} in continuous time.   We assume that a metaorder with participation rate $\eta$ is executed incrementally within $t \in [0,T]$\footnote{Here we do not distinguish volume and physical time.}. The total traded quantity is $Q= \eta T$ and the instantaneous trading rate is $q(t)=-\dot x(t)$, where $x(t)$ is the metaorder quantity that remains to be traded. The price dynamics is
\begin{equation}
S(t)=S(0)+a\int_{0}^{t} q(s) ds + \sigma\int_{0}^{t} \mathrm{d}W_s \ ,
\end{equation}
where $W_t$ is a Wiener process. 
Due to the linearity of the impact function, the immediate impact as a function of time is
\begin{equation}
{\cal I}(t|\Omega = \{ \eta,T \})= a (Q-x(t)) = a (\eta T-x(t))  \ .
\end{equation}
Assuming that during a buy metaorder the trader only buys and never sells, ${\cal I}(t|\Omega = \{ \eta ,T \})$ is a non decreasing function of time converging to the temporary impact $\mathcal{I}_{tmp}(\Omega = \{ \eta, T \})={\cal I}(T|\Omega = \{ \eta,T \})= aQ = a \eta T $, independently of the trading profile followed during the execution. Thus the temporary market impact is a linear function of metaorder duration $T$, for fixed participation rate $\eta$ (see the red solid line in Figure \ref{fig_ac_0}).

What does the immediate impact look like under the Almgen-Chriss model? Even though the impact is linear, if the optimised execution schedule is risk adverse, the immediate impact trajectory {\it during} the execution does not necessarily overlap with the curve described by the temporary impact as a function of $T$ (see the blue lines  in Figure \ref{fig_ac_0}). As a concrete example, consider the optimal trading profile of the original Almgren-Chriss model for a generically risk averse investor. In this case the function to be minimized is the expected cost plus $\lambda$ ($>0$) times the variance of the cost.  The optimal solution is \cite{almgren2001optimal}
\begin{equation}
x(t)=Q\frac{\sinh k(T-t)}{\sinh kT}
\end{equation}
where $k=\sqrt{\lambda \sigma^2/a}$, and $\sigma$ is the volatility of the price. For a risk neutral strategy ($\lambda=0$), the solution is a strategy with constant velocity, $q(s)=\eta$. This is the simplest (and probably more widespread) execution strategy, the so called Volume Weighted Average Price (VWAP) scheme. More risk averse strategies correspond to higher $\lambda$ and lead to more front loaded executions. We observe in Figure \ref{fig_ac_0} that for higher levels of risk aversion the immediate impact deviates more from the temporary impact. In all cases the price reaches the temporary impact from above. This is due to the front loading property of the strategy. For a strategy where the trading rate increases during execution, the price would reach the temporary impact from below. Only in the case of VWAP (i.e. risk neutral strategy) do immediate and temporary impact overlap.

\subsection{The propagator model} \label{sec_prop}

A more sophisticated model is the propagator model, devised to take into account the non linear and immediate properties of market impact. The propagator model was initially proposed by Bouchaud et al. \cite{bouchaud2004fluctuations} in (discrete) transaction time and independently introduced by Lillo and Farmer \cite{lillo2004long} (the latter as a model where price moves in response to the unexpected component of the order flow, see also Taranto et al. \cite{taranto2014adaptive}). An interesting extension that goes beyond the propagator model was very recently proposed in \cite{Donier14}. 

The continuous-time version of the propagator model, discussed by Gatheral \cite{gatheral2010no}, is
\begin{equation}
S(t) = S(0) + \int_{0}^{t} f(q(s)) G(t-s) \mathrm{d}s +\int_{0}^{t} \sigma(s)\mathrm{d}W_s,
\end{equation}
where $G(t)$ is a decaying function describing the temporal dependence of the impact and the function $f(q)$ is an odd function describing the volume dependence of the impact. The Almgren and Chriss model can be recovered by setting $f(q)= aq$ and $G(t)={\mathbb 1}_{t\ge0}$. We consider here a small variation of the Gatheral model where $s(t):= \log S(t)/ \sigma_D$ evolves in volume time $v$ according to
\begin{equation}\label{eq:prop2}
 s(v) =  s(0) + \int_{0}^{v} f(q(s)) G(v-s) \mathrm{d}s +\int_{0}^{v} \mathrm{d} W_s.
\end{equation}
In order to deal with nondimensional quantities, we rescale the instantaneous trading rate $q(s)$ by the daily traded volume $V(t_c)$. Specifically, we consider the propagator model with power-law impact function $f(q) = q^{\delta}$ and power-law decay kernel $G(t) = t^{-\gamma}$. For $\delta = 1$ and $\gamma = 0$ one recovers the Almgren-Chriss model.
Using this model, Gatheral \cite{gatheral2010no} shows that the condition $\delta + \gamma \ge 1$ is necessary to exclude price manipulation. In the case of linear market impact, $f(q)=q$, Gatheral et al. \cite{gatheral2012transient} obtained the optimal condition and derived the explicit form of the optimal strategy in a expected cost minimisation problem. In the general case of non linear impact the problem is more involved \cite{dang2012optimal,curato14}.  In this paper we are not interested in solving the optimisation problem but rather in calculating the market impact for different classes of trading strategies. 
\\

For the simple VWAP strategy characterised by trading rate $q(s)=\eta$ and duration $F$, the temporary market impact is
\begin{equation} 
\mathcal{I}_{tmp}(\Omega=\{ VWAP, \eta, F\}) : =  \int_{0}^{F} f(q(s)) G(F-s) \mathrm{d}s = f(\eta) \int_0^F G(F-s) ds.
\end{equation}
The factorization between the temporal and the volume dependence hypothesized in the propagator model immediately leads to a factorization of the temporary impact into a part depending only on $\eta$ and a part depending only on $F$. In the special case $f(q) = q^{\delta}$ and $G(t) = t^{-\gamma}$ this becomes
\begin{equation} \label{eq_imp_vwap_1}
\mathcal{I}_{tmp}(\Omega=\{ VWAP, \eta, F\}) = \frac{1}{1-\gamma} \eta^{\delta} F^{1-\gamma}. 
\end{equation}
 This relation shows that within the propagator model with a power-law impact function and a power-law decay kernel, the temporary impact is a factorisable power-law function of both the participation rate $\eta$ and of the duration $F$. The ``macroscopic'' exponents describing the shape of the temporary market impact surface $\mathcal{I}_{tmp}(\Omega=\{ VWAP, \eta, F\})$ are inherited from the ``microscopic'' exponents describing the market impact function of individual trades $\delta$ and the decaying kernel $\gamma$. If the model is at the critical condition according to Gatheral, $\delta+\gamma=1$, the temporary impact does not depend on the duration $F$ and the participation rate $\eta$ separately, but only on their product, i.e. the daily fraction $\pi=\eta \cdot F$.  Hence the market impact surface is fully characterized by the market impact curve. 
Notice that with $\gamma=\delta=1/2$, one recovers the square-root impact curve, Eq. \ref{eq_rms}, with $Y=2$. 

Finally, the immediate and the permanent market impact, as defined by Eq. \ref{eq_imp_v}, are

\begin{equation}\label{eq_imp_vwap_2}
\mathcal{I}(z | \Omega=\{ VWAP, \eta, F\})  =
\begin{cases}
\frac{1}{1-\gamma} \eta^{\delta} F^{1-\gamma} z^{1-\gamma} & \text{if } z < 1,\\
\frac{1}{1-\gamma} \eta^{\delta} F^{1-\gamma} \left( z^{1-\gamma} - (z-1)^{1-\gamma} \right)& \text{if } z>1.
\end{cases}
\end{equation}
where $z = v/F$. For large values of $z$ the impact decays to zero as a power law function with exponent $\gamma$. On the other hand, immediately after the metaorder completion, the price decay follows a power law with exponent $1-\gamma$.


The calculation of the immediate and temporary market impact becomes more involved if we consider general execution schemes where the trading velocity is not constant. For purely illustrative purposes, we consider a class of execution schemes of $Q$ shares characterised by the  instantaneous monotonic trading rate
\begin{equation}\label{example}
q(s)=\frac{Q}{V(t_c)}\frac{(\alpha+1)}{F^{\alpha+1}}(F-s)^\alpha,~~~~~~~~~~\alpha>-1.
\end{equation}
For positive (negative) $\alpha$ the relative trading profile trades more (less) at the beginning of the period, while $\alpha=0$ corresponds to a VWAP scheme. It is possible to show that for this class of schemes the temporary market impact is
\begin{equation} \label{tempimpprop}
\mathcal{I}_{tmp}(\Omega=\{ \alpha, \eta, F\}) = \eta^\delta F^{1-\gamma} \frac{(1+\alpha)^\delta}{1+\alpha\delta-\gamma},
\end{equation}
while the immediate and permanent market impact are

\begin{eqnarray}
\mathcal{I}(z | \Omega=\{ \alpha, \eta, F\}) = \eta^{\delta} F^{1-\gamma}(\alpha +1)^{\delta} \int_0^z \mathrm{d}s \frac{(1-s)^{\alpha \delta}}{(z-s)^{\gamma}} {\mathbb 1}_{s\le 1} =  \nonumber \\
\begin{cases}
\frac{1}{1-\gamma} \eta^\delta (1+\alpha)^\delta F^{1-\gamma}  z^{1-\gamma}  ~~ _2\mathcal{F}_1(1,-\delta\alpha;2-\gamma;z) & \text{if } z < 1,\\
\frac{1}{1+\alpha \delta} \eta^\delta (1+\alpha)^\delta F^{1-\gamma}  z^{-\gamma}  ~~ _2\mathcal{F}_1(1,\gamma;2+\delta\alpha;\frac{1}{z}) & \text{if } z>1,
\end{cases}
\end{eqnarray}
where $z=v/F$ is the rescaled time and $_2\mathcal{F}_1$ is the hypergeometric function. In order to avoid the divergence of the temporary impact ($z=1$), one has to impose $1+\alpha\delta-\gamma>0$. 
\\

\begin{figure}[t] 
  \centering
	\includegraphics[width=1\textwidth]{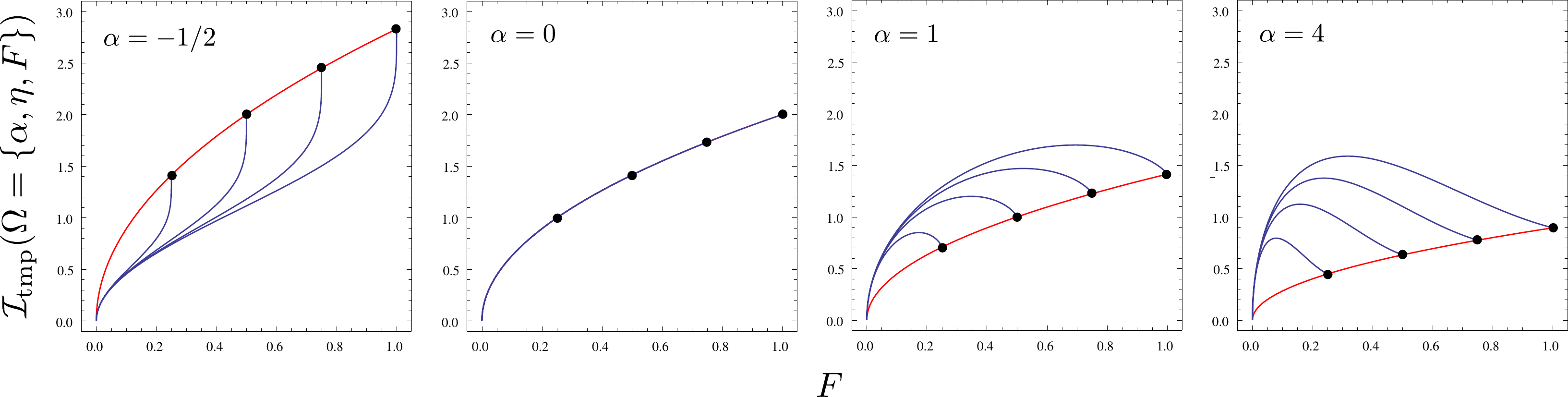}
  \caption{ Temporary market impact (red line) $\mathcal{I}_{tmp}(\Omega=\{ \alpha, \eta, F\})$ as a function of the metaorder duration $F$ for fixed participation rate $\eta=1$ and different values of trading rate profiles $\alpha$ for the propagator model with $\gamma=\delta=1/2$. We consider a back loaded profile ($\alpha = -1/2$, top left), a VWAP profile ($\alpha = 0$, top right), and two front loaded profiles ($\alpha = 1$ and $\alpha = 4$, bottom). In each panel we also show the immediate market impact trajectory (blue lines) $\mathcal{I}(v | \Omega=\{ \alpha, \eta, F\})$ for some values of $F=0.25,0.5,0.75,1$ as a function of volume time $v$.
  }
  \label{fig_ac}
\end{figure}

Figure \ref{fig_ac} shows the temporary market impact $\mathcal{I}_{tmp}(\Omega=\{ \alpha, \eta, F\})$ as a function of the metaorder duration $F$ at fixed participation rate $\eta$ for different values of trading profiles $\alpha$ in a propagator model with $\gamma=\delta=1/2$. For some values of the duration $F$ we also show the immediate market impact trajectory $\mathcal{I}(v | \Omega=\{ \alpha, \eta, F\})$, i.e. the trajectories followed by the price to reach the temporary impact. There are three important comments one can make observing these figures: (i) The temporary market impact depends on the trading profile (see also Eq. \ref{tempimpprop}). This is a general consequence of the non-linearity of the impact function $f$. As we have seen before in the Almgren-Chriss model, the temporary impact is independent of the trading profile. (ii) As in the Almgren-Chriss model, the more the trading profile deviates from VWAP, the more the immediate impact trajectories deviate from the temporary impact. For front (back) loaded strategies, $\alpha>0$ ($\alpha<0$), the trajectories reach the temporary impact from above (below); (iii) Even if the trade sign is always the same during execution (e.g. buys for a buy metaorder), the impact trajectories can be non-monotone, since they reach a maximum and decay to the temporary impact, {\it before} the end of the metaorder (see the cases $\alpha=1$ and $\alpha=4$ in Fig. \ref{fig_ac}). In other words, the price reversion, well documented after the end of the metaorder, starts during the metaorder's execution if the trading profile is front loaded enough. As we will see below, this is exactly what we observe for real metaorder executions in some of the data.


\section{Measurements of market impact}\label{sec:measurement}

\subsection{Market impact curve: testing the square-root formula}

\begin{figure}[t] 
  \centering
	\includegraphics[width=0.8\textwidth]{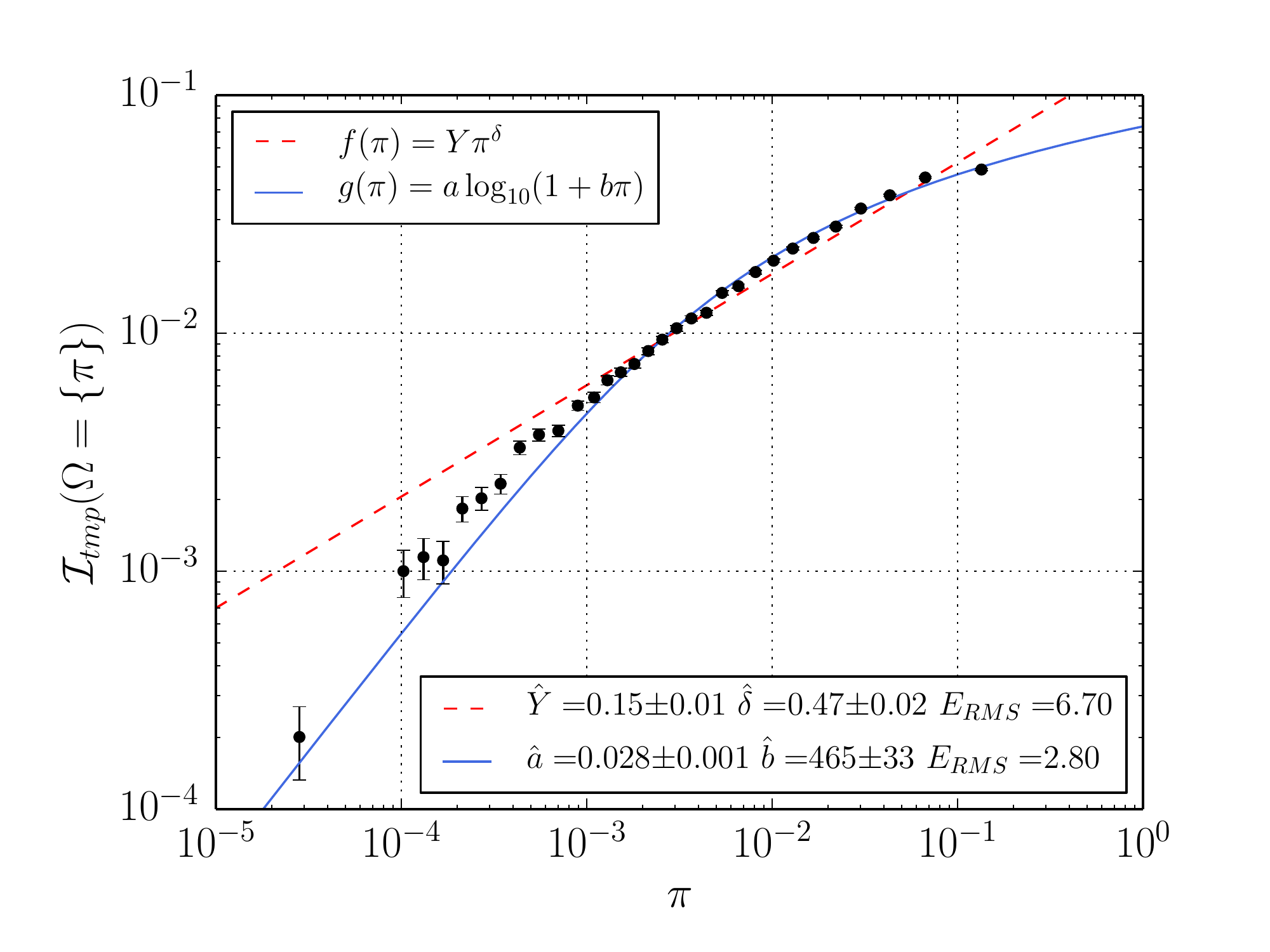}
  \caption{Measured temporary market impact $\mathcal{I}_{tmp}(\Omega=\{ \pi \} )$ of a metaorder as a function of the daily rate $\pi$, defined as the ratio of the traded volume and the daily volume. The scale is double logarithmic; the dashed read line is the best fit to a power-law and the solid blue curve is the best fit to a logarithm.}
  \label{fig_toth0}
\end{figure}

The temporary market impact curve $\mathcal{I}_{tmp}(\Omega=\{ \pi \} )$ is defined as the price change conditioned on the daily fraction $\pi$. The square-root impact formula (Eq. \ref{eq_rms}) states that the temporary market impact curve is described, at least to a first approximation, by a power-law function:
\begin{equation}\label{eq_rms2}
\mathcal{I}_{tmp}(\Omega=\{ \pi \}) =Y \pi^\delta \ . 
\end{equation}
Previous studies find $\delta$ in the range 0.4 to 0.7 \cite{almgren2005direct,toth2011anomalous, mastromatteo2014agent, brokmann2014}. Figure \ref{fig_toth0} shows the shape of the measured temporary market impact curve for our data in double logarithmic scale. This plot is obtained by dividing the data into evenly populated bins according to the daily rate $\pi$ and computing the conditional expectation of the impact for each bin. Here and in the other figures of this paper, the error bars are standard errors. Note that the range of $\pi$ spans more than five orders of magnitude. We observe that for $\pi$ roughly in the range from $10^{-3}$ to $10^{-1}$ the points, to a first approximation, lie on a straight line.  Nonetheless, a clear concavity is evident, since for large and small $\pi$ the impact curve bends down.

Performing a nonlinear regression on the function $f(\pi|Y,\delta) = Y \pi^{\delta}$, the best fitting parameters are $\hat Y  = 0.15 \pm 0.01$ and $\hat\delta=0.47\pm0.01$. The value of the exponent is consistent with that found in previous work \cite{almgren2005direct,toth2011anomalous, mastromatteo2014agent, brokmann2014}. 
In order to compare different functional forms, we consider the Weighted Root Mean Square Error, $E_{RMS}$, as a measure of the goodness of fit\footnote{The Weighted Root Mean Square Error of a function $g(x|a,b)$ with parameters $a$ and $b$ to reproduce the observations $\{ y_i \}_{1 \le i \le N}$ of the explanatory variables $\{ x_i\}_{1 \le i \le N}$ is defined by \begin{equation}
E_{RMS}(g(a, b)) = \sqrt{ \frac{1}{N}\sum_i^N \left( \frac{y_i-g(x_i| a, b) }{SE(y_i)} \right)^2 } 
\end{equation} where $SE(y_i)$ is the standard error associated with the observation $y_i$. The smaller $E_{RMS}$, the better the fit.}. For the power law we find $E_{RMS}(f(\hat{Y}, \hat{\delta})) = 6.70$. The concavity of the market impact shape depicted in double-logarithmic scale suggests that a function more concave than a the power-law might better explain the data. As an alternative we fit a function of the form $g(\pi|a,b) = a \log_{10}(1+b\pi)$. The shape of $g(\pi|a,b)$ is linear for values of $b\pi \ll 1$ and logarithmically concave for $b\pi > 1$. The estimated best fitting parameters are $\hat a  = 0.028 \pm 0.001$ and $\hat b=465 \pm 33$ and the relative goodness of fit and $E_{RMS}(g(\hat{a}, \hat{b})) = 2.80$, which is quite dramatically better than that for the power law.   The light blue line in Figure \ref{fig_toth0} is the best fitting logarithmic curve. From the figure and from the values of $E_{RMS}$, we conclude that the logarithmic functional form describes our data better than the power-law (square root) functional form.  
\\

\begin{figure}[t] 
  \centering
	\includegraphics[width=0.38\textwidth]{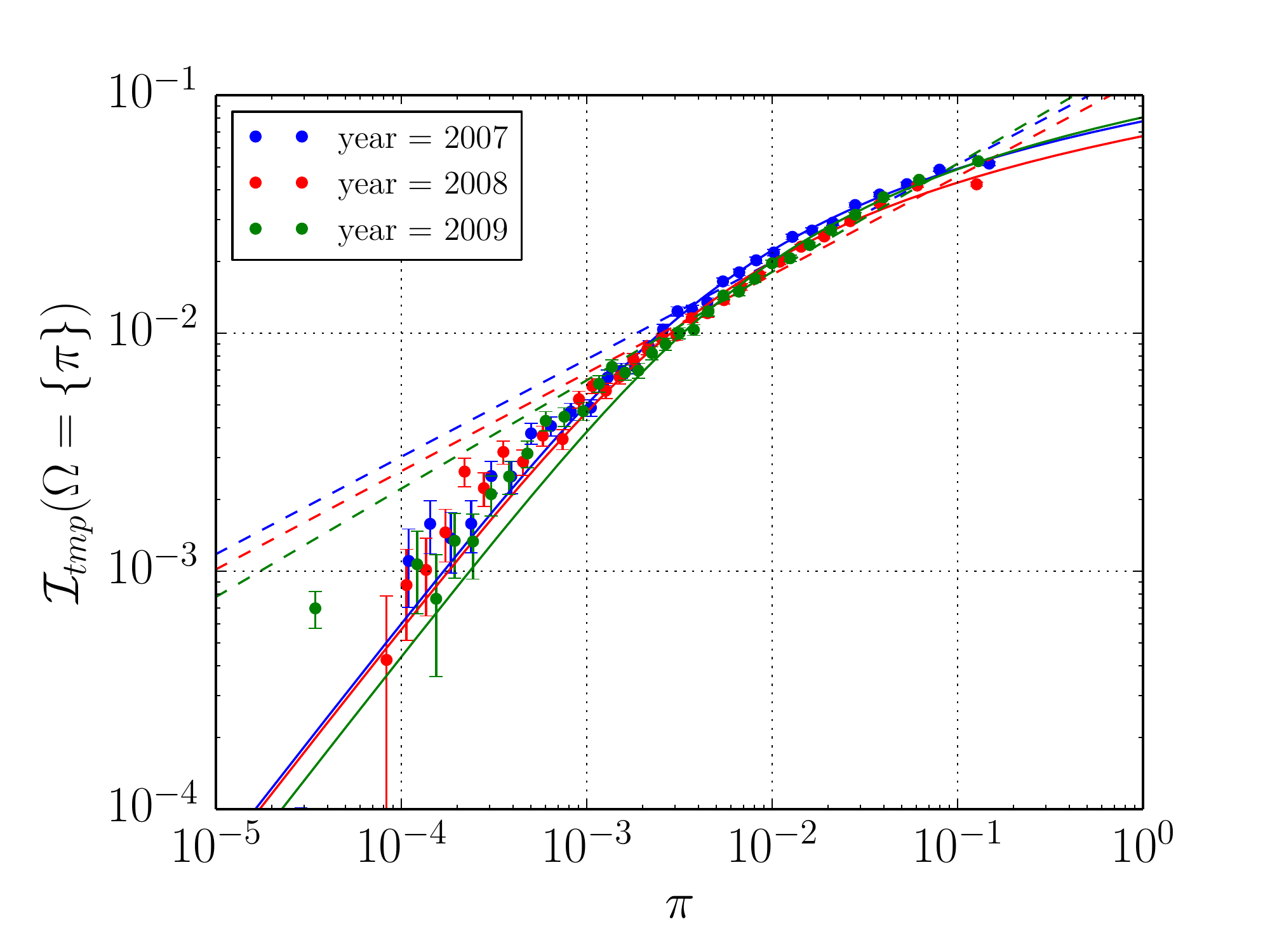}%
	\includegraphics[width=0.38\textwidth]{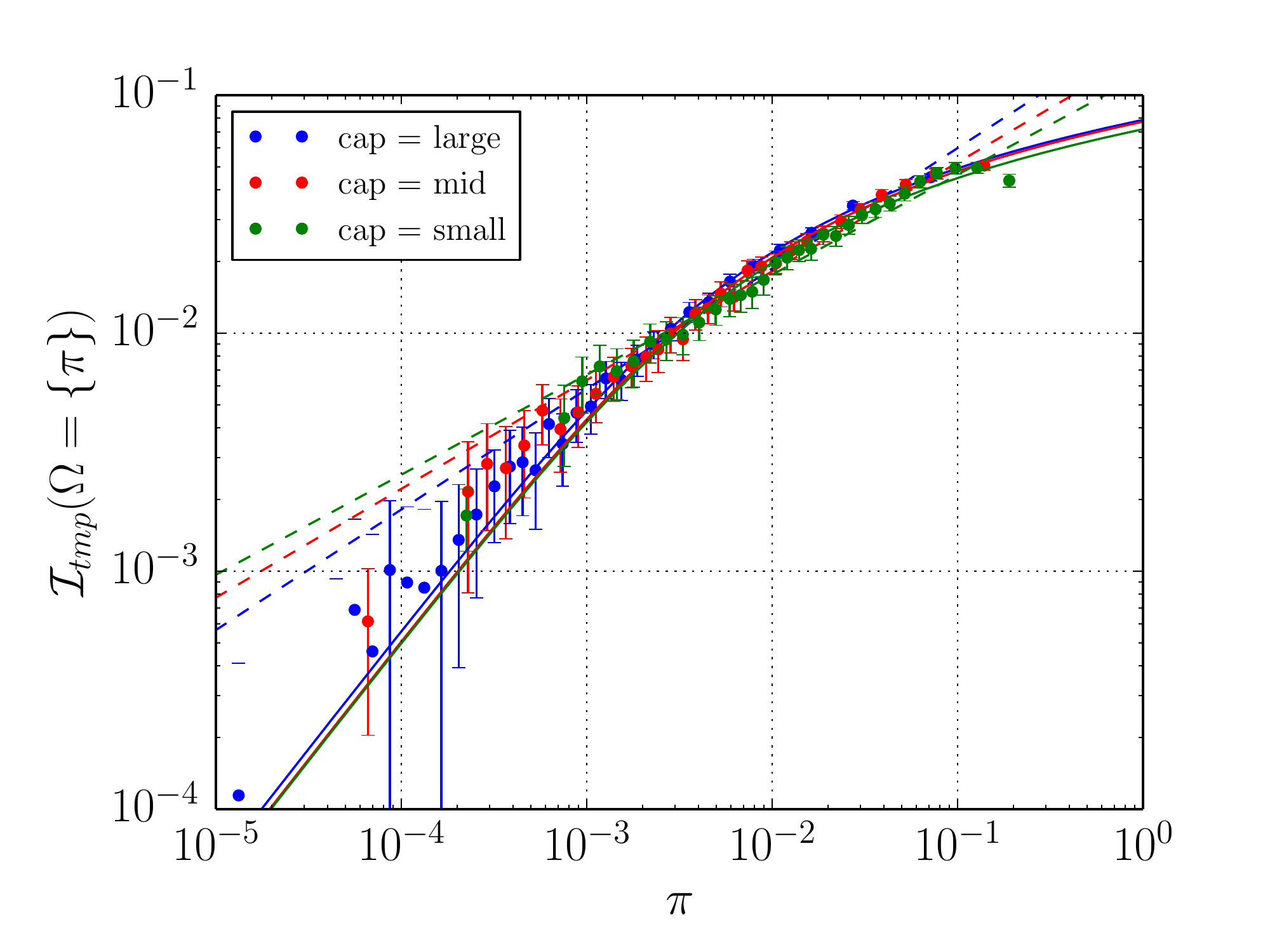}\\
     	\includegraphics[width=0.38\textwidth]{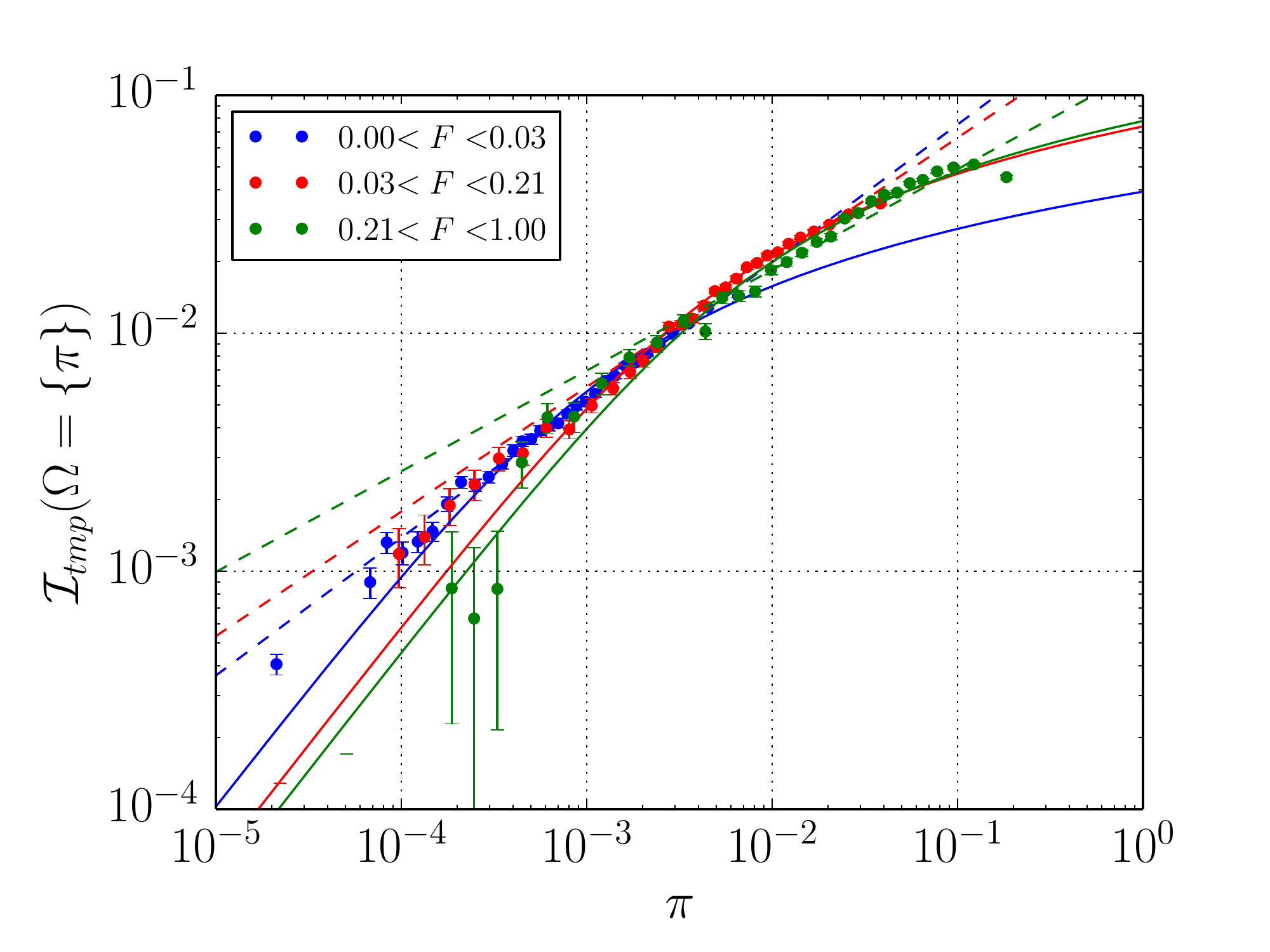}%
	\includegraphics[width=0.38\textwidth]{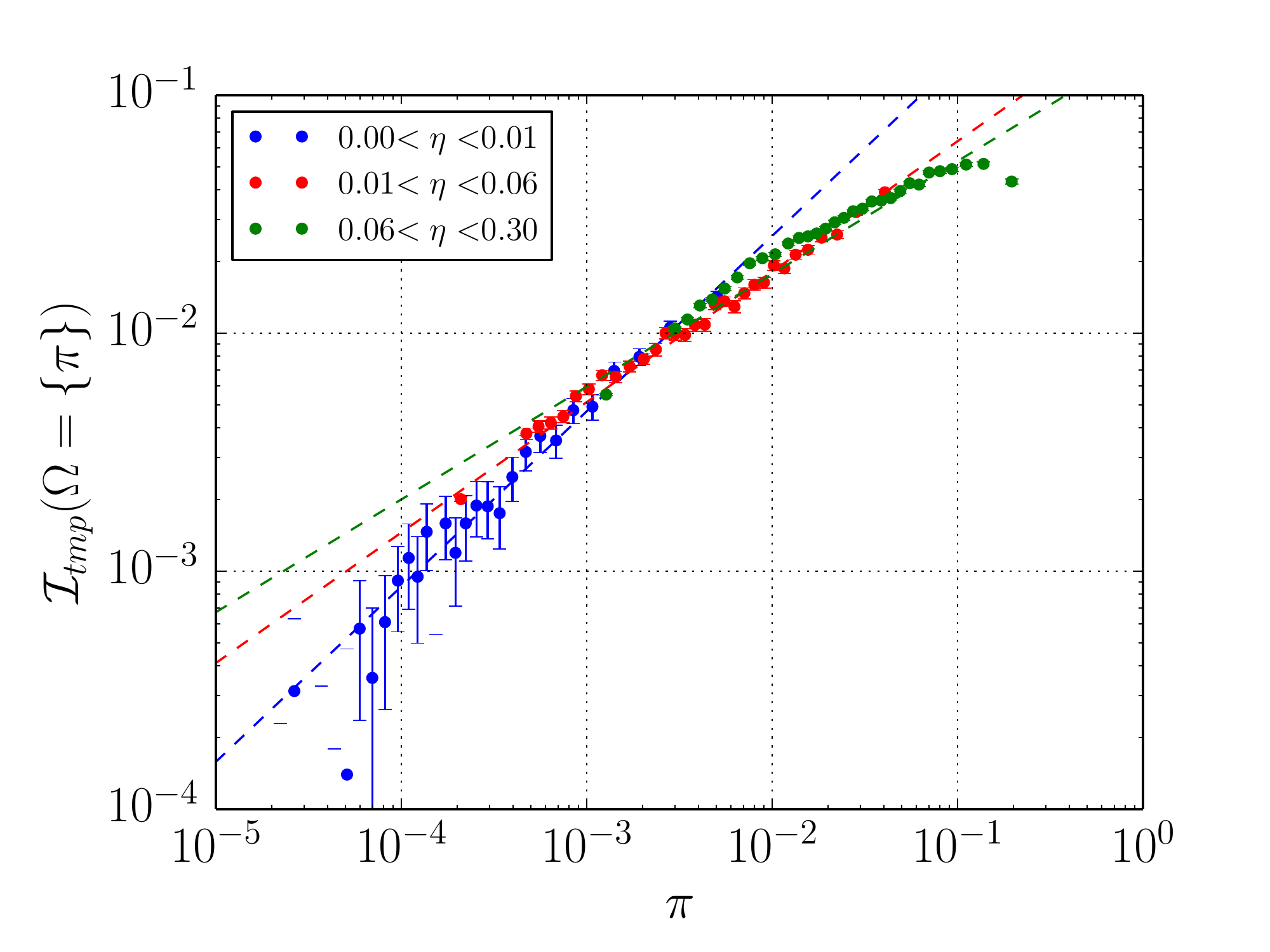}
  \caption{Temporary market impact of a metaorder as a function of the daily fraction $\pi$, defined as the ratio of the traded volume and the daily volume. The scale is double logarithmic and the lines are a best fit with a power-law function $f(\pi| Y, \delta)$ (dashed) and a logarithmic function $g(\pi | a,b)$ (solid). The top left panel considers three different years separately, the top right panel considers the market capitalisation of the traded stock separately, the bottom left panel considers different subsets of the metaorder duration $F$, and the bottom right panel considers different subsets of the participation rate $\eta$.  The values of the best fitting parameters and the goodness of fit for the power law and the logarithmic function are reported in table \ref{tab_fit} for each year and market capitalisation. }
  \label{fig_toth}
\end{figure}

We now consider the problem of how the temporary impact curve depends on other conditioning variables.
As suggested in \cite{toth2011anomalous}, the square-root law of market impact seems to be a very robust statistical regularity. It does not appear to depend on the traded instrument (equities, futures, FX, etc.) or time period (from the mid-nineties, when liquidity was provided by market makers, to present day electronic markets). We verify the robustness of the temporary market impact curve by conditioning on different time periods present in the database and on the market capitalisation of the traded instrument. Figure \ref{fig_toth} (top panels) shows the results of our analysis: quite remarkably, the shape of the temporary market impact curve is roughly independent on both the trading period (2007, 2008 and 2009) and the stock capitalisation (small, medium and large, according to the classification provided by the Ancerno database). The estimated best fitting parameters and the relative goodness of fit are reported in table \ref{tab_fit}. We observe that, with the exception of the small capitalisation conditioning, the logarithmic function is always better in explaining the data.

We now shift our focus to measuring the temporary market impact curve while conditioning on the parameters characterising the execution of a metaorder. As pointed out in \cite{gatheral2010no}, the square-root formula depends only on the daily fraction $\pi$, which implies that the temporary market impact is independent of both the duration $F$ and the participation rate $\eta$. We investigate this point by conditioning our measurements on the participation rate and duration of the metaorder. 
The bottom panels of Figure \ref{fig_toth} show the results of this analysis  We notice that in both cases the three curves are locally approximated by a power law function, essentially because the two types of conditioning reduce significantly the span of data on the abscissa. However it is clear that the exponents of the power law are different and, as a consequence, the superposition of the three subsets gives a logarithmically concave function. In particular the power law exponent decreases from $0.58$ to $0.42$ when we condition impact on $F$, and from $0.74$ to $0.47$ when we condition on $\eta$. In all cases the error on the expoenent is $0.02$. Therefore the exponent, and thus the impact function, depends on the conditioning variable
($F$ or $\eta$).

\begin{table}[t]
\begin{center}
\begin{tabular}{| l | l | r | r | r |}
\hline
 & Year & 2007 & 2008 & 2009 \\
  \hline                       
Power law&$\hat{Y}$  & 0.13$\pm$0.01 & 0.12$\pm$0.01 & 0.15$\pm$0.01 \\
&$\hat{\delta} $ & 0.41$\pm$0.02  & 0.41$\pm$0.02 & 0.46$\pm$0.01\\
&$E_{RMS} $ & 5.47 & 4.96 & 3.18\\
\hline
Logarithm&$\hat{a}  $& 0.029$\pm$0.001 & 0.025$\pm$0.001 & 0.032$\pm$0.001\\
&$\hat{b}  $& 491$\pm$29 & 547$\pm$56 & 316$\pm$25\\
&$E_{RMS}  $& {\bf 1.36} & {\bf 2.11} & {\bf 2.23}\\
  \hline  
  \hline
& Mkt. Cap & Large & Medium & Small \\
  \hline                       
Power Law&$\hat{Y}$  & 0.19$\pm$0.01 & 0.15$\pm$0.01 & 0.12$\pm$0.01 \\
&$\hat{\delta} $ & 0.51$\pm$0.02 & 0.46$\pm$0.02 & 0.42$\pm$0.02 \\
&$E_{RMS} $ & 1.38 & 1.32 & {\bf 0.69}\\
\hline
Logarithm &$\hat{a}  $ & 0.030$\pm$0.001 & 0.030$\pm$0.001 & 0.027$\pm$0.001\\
&$\hat{b}  $ & 441$\pm$21 & 400$\pm$35 & 428$\pm$62 \\
&$E_{RMS}  $ & {\bf 0.40} & {\bf 0.69} & 1.03 \\
  \hline  
\end{tabular}
\end{center}
\caption{Estimated values of the best fitting parameters $\hat{Y}$ and $\hat{\delta}$ for the fitting function $f(\pi | Y, \delta) = Y \pi^{\delta}$ and $\hat{a}$ and $\hat{b}$ for the fitting function $g(\pi | a,b) = a \log_{10}(1+b \pi )$ and the corresponding goodness of fit. Each fit is performed conditioning the sample data on the execution year (top table) and market capitalisation (bottom table). In bold face there is the fits with the smallest $E_{RMS}$.}
\label{tab_fit}
\end{table}

In conclusion, when plotted as a function of the daily rate $\pi$, the temporary market impact curve $\mathcal{I}_{tmp}(\Omega=\{ \pi \})$ is clearly described by a concave function, well fitted by a logarithmic function and only locally approximated by a square root function. Interestingly strong concavity for very large volumes has been quoted for CFM metaorders also in reference \cite{Donier14}. In the next section we show that the impact does depend on $F$ and $\eta$ separately, and that the collapse seen here is significantly due to a compensation effect from the data aggregation of orders with different conditioning parameters.


\subsubsection{Inferring latent order book from market impact\label{virtualOrderBook}}

In a recent paper Toth et al. \cite{toth2011anomalous} present a theory connecting the shape of the market impact to the one of latent order book.  They argue that true order book does not reflect the actual supply and demand that are present in the market, due to the fact that participants do not reveal their true intensions. Latent order book becomes visible when price moves and thus can be inferred from the shape of market impact.

Specifically, let ${\cal V}(x|b,n)$ be the volume available in the latent order book at log price $x$.  A metaorder of size $\pi$ will generate a market impact ${\cal I}(\pi)$ solving the equation
\begin{equation}\label{eqlob}
\pi=\int_{x_0}^{x_0+{\cal I}(\pi)} {\cal V}(x|b,n)~~dx
\end{equation}
 where $x_0$ is the log price at the beginning of the metaorder execution.
If the profile ${\cal V}$ is a linear function of log price, then market impact is a square root function of the traded volume $\pi$, as suggested by \cite{toth2011anomalous}. On the other hand, if the profile is constant, the market impact is linear in the traded volume. 

Here we assume a parametric form, and which for the center of the order book allows us to easily interpolate between a linear vs. a constant order book.  We then compute the expected  market impact and we fit the parameters of the latent order book profile on real data of market impact. Specifically, we consider the normalized function 
\begin{equation}
{\cal V}(x|b,n):= \frac{1}{Y}\frac{ x^n \exp(bx)}{\int_0^1 \mathrm{d} y \  y^n \exp(by)}
\label{bookProfile}
\end{equation}
where $x \in [0,1]$, ${\cal V}(0|b,n)=0$ and ${\cal V}(1|b,n)=1/Y$, where $Y$ is a normalizing constant.  For $x \ll 1/b$ the profile grows as a polynomial, while for $x \gg 1/b$ it grows as an exponential. We can invert Eq. \ref{eqlob} to derive $\mathcal{I}(\pi |Y,b,n)$ obtaining the following cases. When $n=0$, we can perform the analytical calculation and recover the previously introduced logarithm function: $\mathcal{I}(\pi |Y,b,n=0) :=Y \log(1 + c\pi )/\log(1+c)$ where $c=\exp(b)-1$. The market impact function grows linearly for $\pi \ll 1/c$ and logarithmically for $\pi \gg 1/c$.
When $n=1$ we obtain a market impact function $\mathcal{I}(\pi |Y,b,n=1)$ that grows as a square root function for small $x$ and logarithmically for large $x$.  By keeping $n$ as a free parameter we can infer the order book shape near the best by fitting the impact function.


Because the data set is so large we divide the data in into $N_{bins}$ evenly populated bins. For each bin $i$ we measure the average daily rate $\pi_i$, the average impact $\mathcal{I}_i$, and the standard error on the sample impact $SE(\mathcal{I}_i)$. Then, via a non-linear weighted optimisation, we obtain the best fitting parameters of the impact models $\mathcal{I}(\pi |Y,b,n=0)$, $\mathcal{I}(\pi |Y,b,n=1)$, and $\mathcal{I}(\pi |Y,b,n)$ and we calculate the Weighted Root Mean Squared Error of each model. The large number of bins in the fitting procedure compared to the number of parameters minimizes the risk of overfitting\footnote{The Akaike Information Criterion leads to the same conclusions on the relative performance of the different functional forms.}.  We have also tested the conclusions by varying the number of bins and find that for $N_{bins} > 500$ the results are independent of the number of bins.

\begin{figure}[t] 
\centering
 	\includegraphics[width=0.8\textwidth]{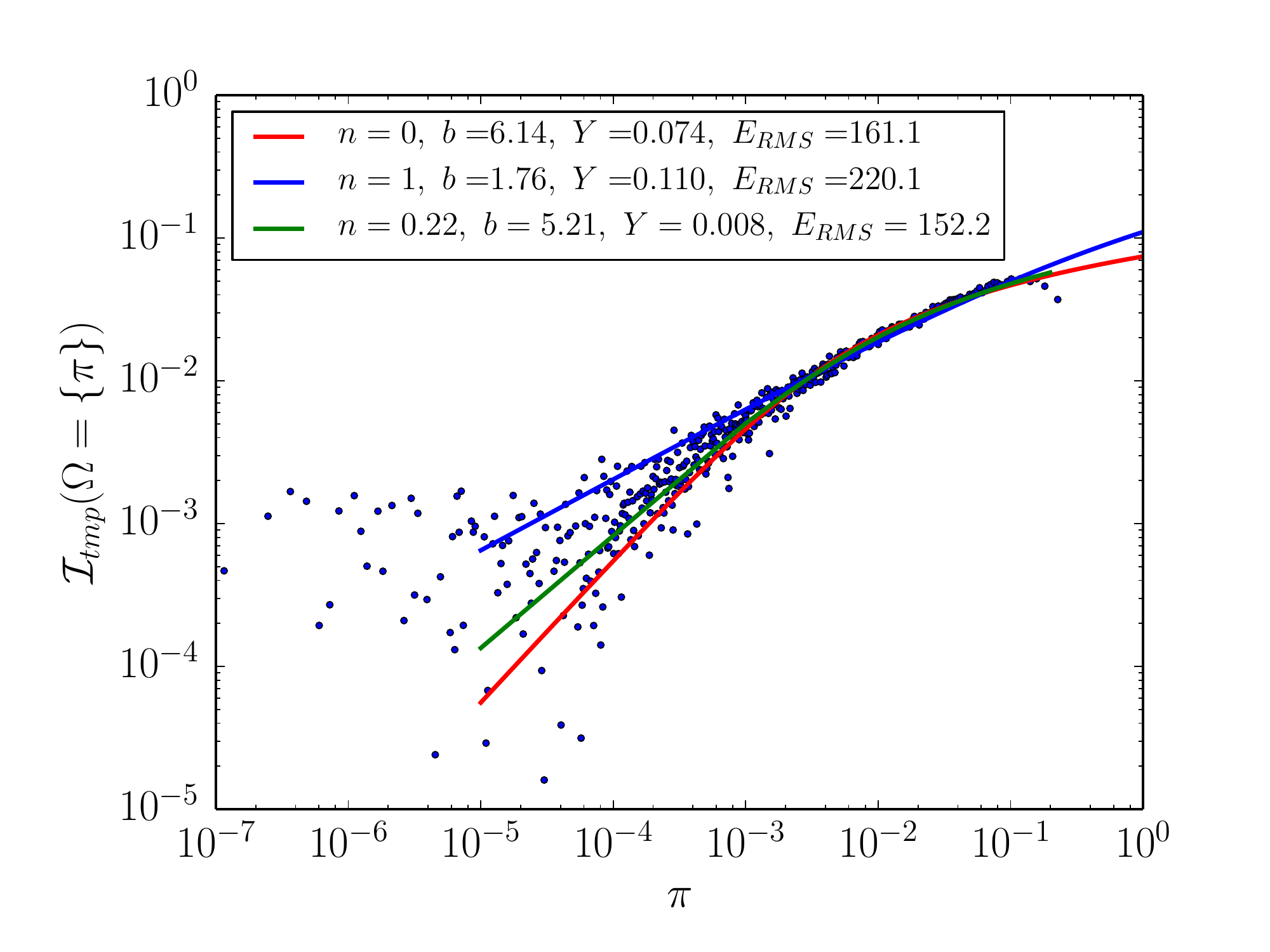}
	\caption{The inferred impact function based on a latent order book profile parameterized with Eq.~(\ref{bookProfile}) and solved using Eq.~(\ref{eqlob}).  The inferred impact is plotted on double logarithmic scale as a function of the normalized order size $\pi$.  See Eq.~(\ref{bookProfile}) for the interpretation of the parameters. }
  \label{fig_lob}
\end{figure}

The results of this procedure are reported in figure \ref{fig_lob}. We observe that the model  $\mathcal{I}(\pi |Y,b,n=0)$ describes the data better than the model $\mathcal{I}(\pi |Y,b,n=1)$. The fitted value of the parameter $1/c$ discriminating the linear from the logarithmic regime in the impact in the case $n=0$ is $\pi^*\simeq 2 \times 10^{-3}$, indicating that when $\pi\ll\pi^*$ market impact is linear, while above this value the impact starts to be logarithmic. The model with three free fitting parameters $\mathcal{I}(\pi |Y,b,n)$ clearly improves the goodness of fit. The value of the inferred exponent, $\hat{n}=0.22$, is close to zero, suggesting an almost flat order book profile near the best bid/ask positions, even if the noise observed in the left part of figure \ref{fig_lob} is quite large. From this we conclude that our data is consistent with an exponential form of the latent order book for large volumes (see also the discussion in Section X of the recent paper \cite{Donier14}).

\subsection{Market impact surface: beyond the square-root formula}

\begin{figure}[t] 
  \centering
	\includegraphics[width=0.95\textwidth]{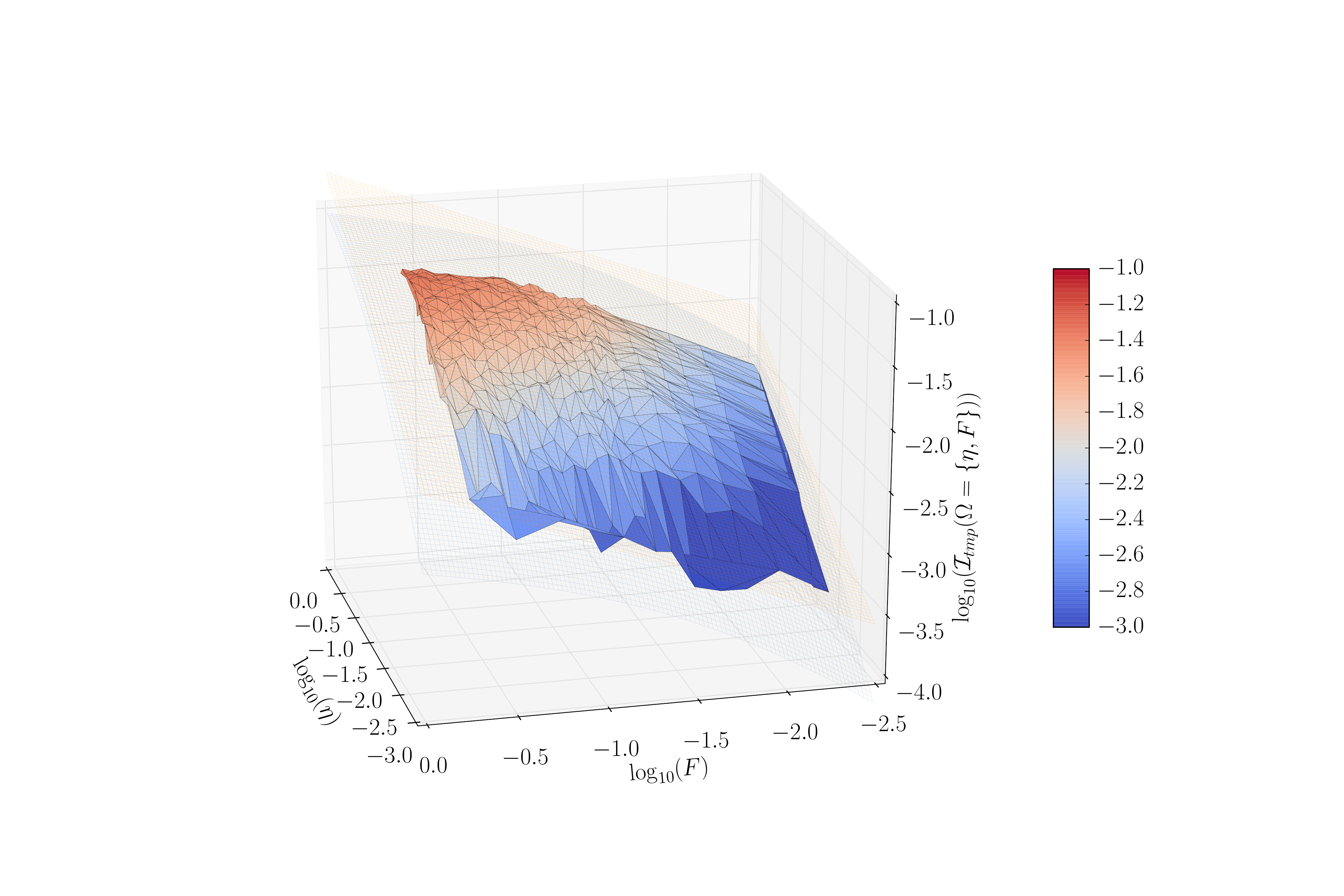}
  \caption{Non-parametric estimation of the impact surface ${\cal I}_{tmp}(\Omega = \{ F,\eta \})$ as a function of the duration $F$ and the participation rate $\eta$. The three axes are in logarithmic scale. The orange surface represents the double power-law function $f(\eta,F|\hat{Y},\hat{\delta},\hat{\gamma_1}) = \hat{Y} \eta^{\hat{\delta}} F^{\hat{\gamma_1}}$ with the empirically fit parameters $\hat{Y}$, $\hat{\delta}$ and $\hat{\gamma_1}$. The blue surface represents the double logarithmic function $g(\eta,F|\hat{a},\hat{b},\hat{c}) = \hat{a} \log_{10}(1+\hat{b} \eta) \cdot \log_{10}(1+\hat{c} F)$ with the empirically fit parameters $\hat{a}$, $\hat{b}$ and $\hat{c}$. }
  \label{fig_toth_3d}
\end{figure}

We now consider the dependence of the temporary market impact  $\mathcal{I}_{tmp}(\Omega=\{ \eta,F \})$ on the participation rate $\eta$ and the duration $F$ of metaorder execution. Please recall that in Figure~\ref{fig_stat} we demonstrated that the correlation between these two parameters is quite low. 

Figure \ref{fig_toth_3d} shows a non parametric estimation of the market impact surface $\mathcal{I}_{tmp}(\Omega=\{ \eta,F \}) $ for the roughly five million metaorders in our database\footnote{In the present analysis we subset for $\eta>10^{-3}$, in order to avoid a strongly noisy part of the domain. We subset also for $F<0.5$, since the remaining part the plot shows the most discrepant behaviour compared to the square root.}. The three axes of the plot are in logarithmic scale to highlight the concave shape of the market impact surface. If the temporary impact is described by a power-law function both in $\eta$ and in $F$, i.e.
\begin{equation}\label{eq_rms3}
\mathcal{I}_{tmp}(\Omega=\{ \eta,F \}) = Y \cdot \eta^{\delta} \cdot F^{\gamma_1} 
\end{equation}
the market impact surface is a plane in logarithmic scale. Figure \ref{fig_toth_3d} shows that a linear functional form is only an approximate representation of the empirical surface. In fact the surface is clearly concave (in log scale), and almost flattens out in the  top left corner. We perform a non-linear regression of the measured temporary market impact surface $\mathcal{I}_{tmp}(\Omega=\{ \eta,F \})$ with a power law function $f(\eta, F|Y,\delta,\gamma_1)= Y \eta^{\delta} F^{\gamma_1}$, according to equation \ref{eq_rms3}. The best fitting parameters are $\hat Y = 0.207 \pm 0.005$,  $\hat \delta=0.52 \pm 0.01$ and  $\hat \gamma_1 = 0.54 \pm 0.01$. The Root Mean Square Error for this model is $E_{RMS}(f(\hat{Y},\hat{\delta},\hat{\gamma_1}))=2.46$. Interestingly these values are very close to those predicted by the critical propagator model with $\delta=\gamma=1/2$. The orange plane in Figure \ref{fig_toth_3d} is the functional form $f(\eta,F|\hat{Y},\hat{\delta},\hat{\gamma_1})$ with the best fitting parameters. 

\begin{figure}[t] 
  \centering
 	\includegraphics[width=0.5\textwidth]{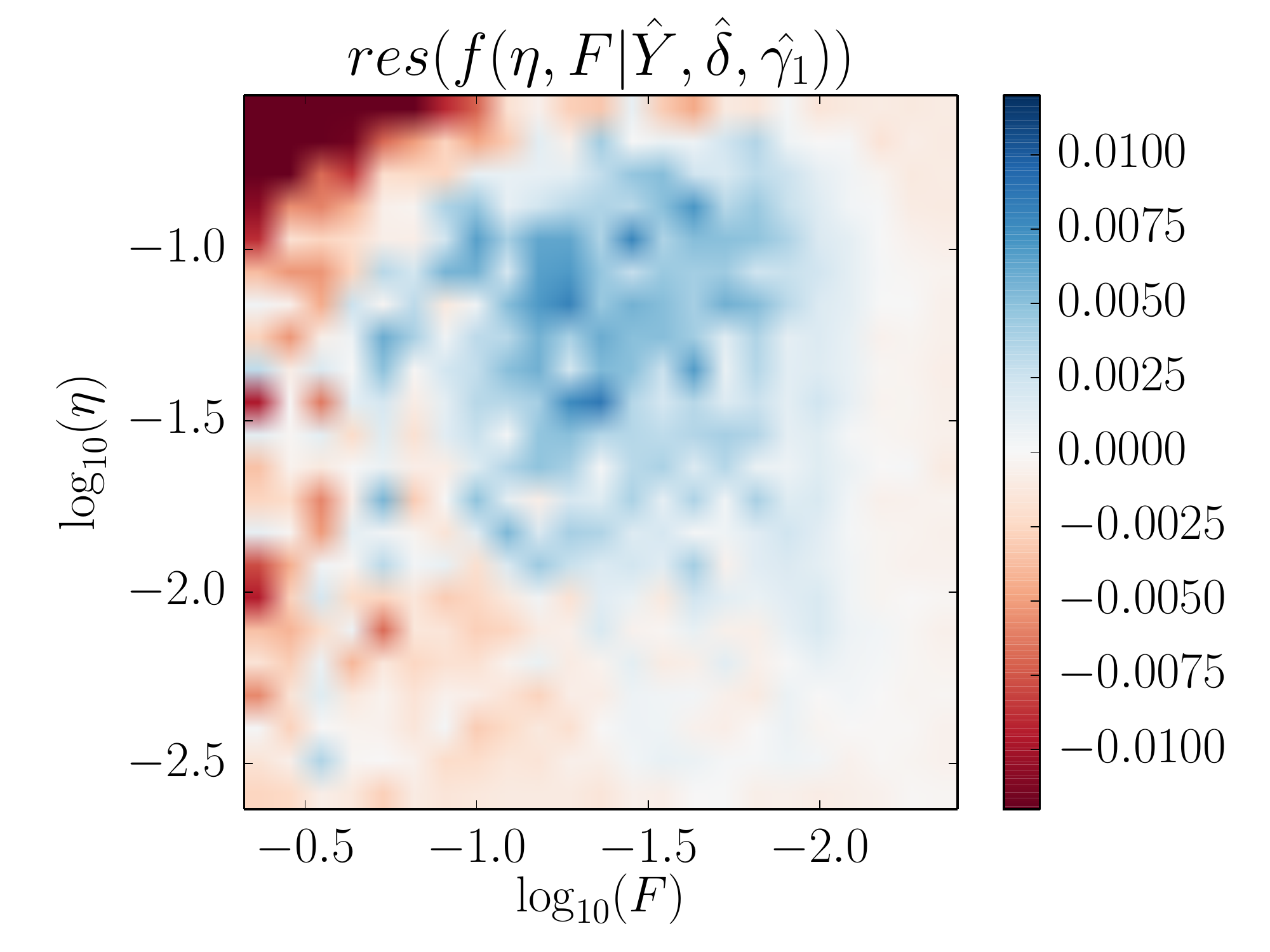}%
	\includegraphics[width=0.5\textwidth]{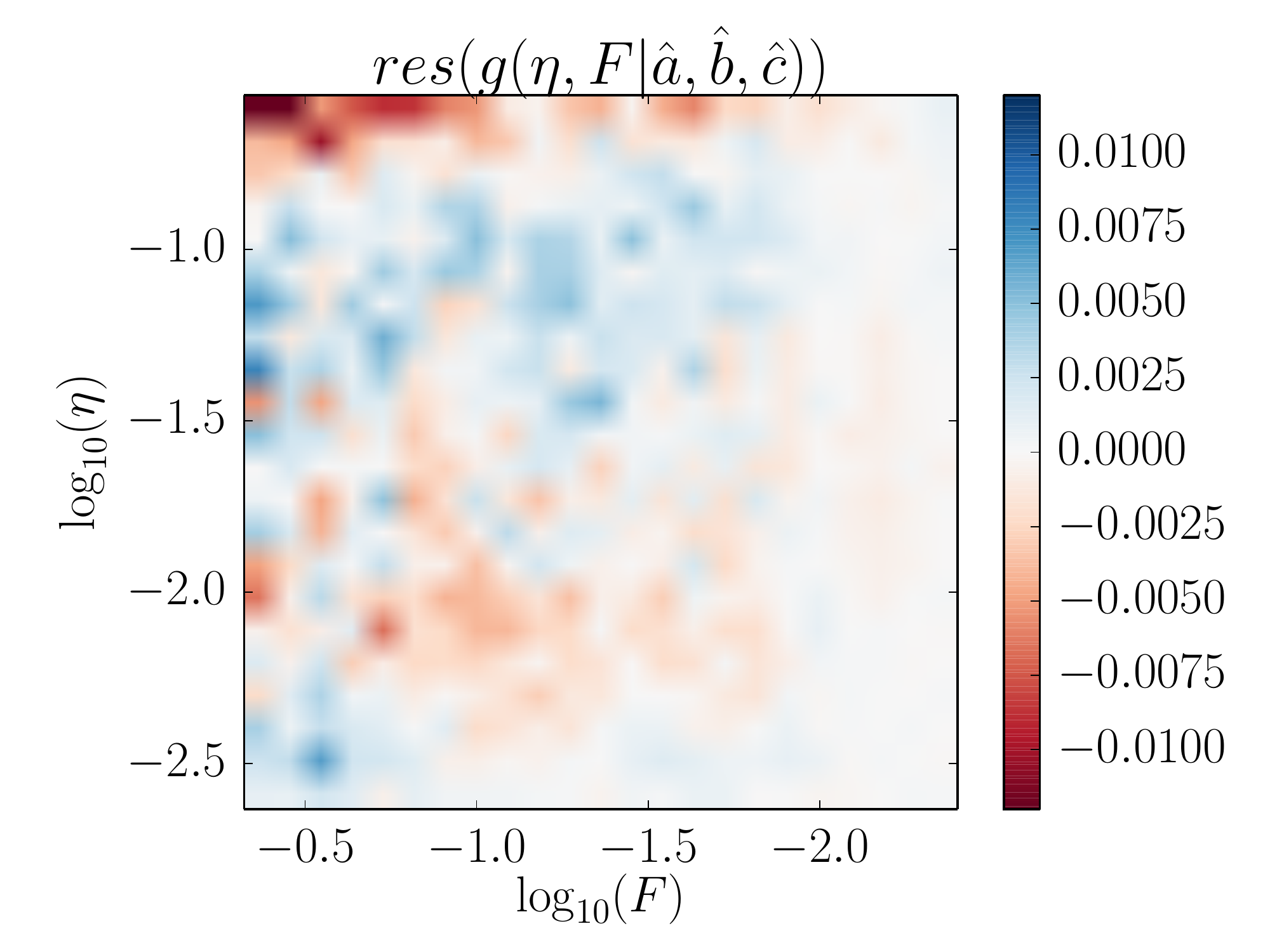}
  \caption{Contour plot of the residuals of the fitting function $f(\eta,F|\hat{Y},\hat{\delta},\hat{\gamma_1}) = \hat{Y} \eta^{\hat{\delta}} F^{\hat{\gamma_1}}$ (left panel) and of the fitting function $g(\eta,F|\hat{a},\hat{b},\hat{c}) = \hat{a} \log_{10}(1+\hat{b} \eta)  \cdot \log_{10}(1+\hat{c} F)$ (right panel) as a function of the plane $\eta$ -- $F$. Positive (negative) residuals are in blue (red).  The power law clearly has macroscopic structure in the residuals; this is much less so for the logarithm.}
  \label{fig_res}
\end{figure}

In order to quantify the deviations of the surface from the best fitting power-law function $f(\eta,F|\hat{Y},\hat{\delta},\hat{\gamma_1})$, in the left panel of Figure \ref{fig_res} we show the residuals of the fit as a function of $\eta$ and $F$. Positive residuals are in blue, while negative residuals are in red. A clear non-random pattern emerges, since residuals in the center are typically positive, while those in the periphery are negative. This is an indication of the approximate description provided by Eq. \ref{eq_rms3} and therefore by the square root law: the impact surface is concave even in logarithmic scale. This suggests that an improvement of the parametrisation of the market impact surface could be obtained considering a logarithmic functional form\footnote{Note that the parameters $a$ are $b$ are not necessarily the same as those used in the parametrisation of the impact curve. The relation between the parameters of the curve and of the surface depends on the joint distribution of $F$ and $\eta$.}  $g(\eta, F | a,b,c) = a \log_{10}(1+b \eta) \log_{10}(1+c F) 
$. By means of a non-linear regression, we obtain the best fitting parameters: $\hat a = 0.035 \pm 0.001$, $\hat b = 60 \pm 3$ and $\hat c = 61 \pm 2$. The Mean Square Error of the fit is $E_{RMS}(g(\hat{a}, \hat{b}, \hat{c})) = 1.44$. This last value is much smaller than the $E_{RMS}$ of the double power law function of Eq. \ref{eq_rms3} and, as in the previous section, indicates that logarithmic functions better describe temporary market impact.  The blue surface in figure \ref{fig_toth_3d} represents the functional form  $g(\eta, F | \hat{a},\hat{b},\hat{c})$ evaluated with the best fitting parameters. In the right panel of figure \ref{fig_res} we present the residuals of the regression $g(\eta, F | \hat{a},\hat{b},\hat{c})$: the pattern present on the left panel for the power law fit is very strongly attenuated, indicating once more a better fit.

The analysis of the residuals also allows us to understand why the square root gives a relatively good collapse of the data, apparently independent from internal and external conditioning variables. In fact, consider the structure of the residuals in the left panel of Figure \ref{fig_toth_3d}. Since $\pi=\eta F$, conditioning on $\pi$ means taking averages over diagonal strips going from the bottom left to the top right part of the plane $(\eta, F)$. This averaging includes positive and negative residuals that partly cancel out, giving the observed data collapse.  Nonetheless, the disaggregation of the data done here indicates that there is indeed dependence on $F$ and $\eta$ separately.
\\


\begin{figure}[t]
  \centering
	\includegraphics[width=0.5\textwidth]{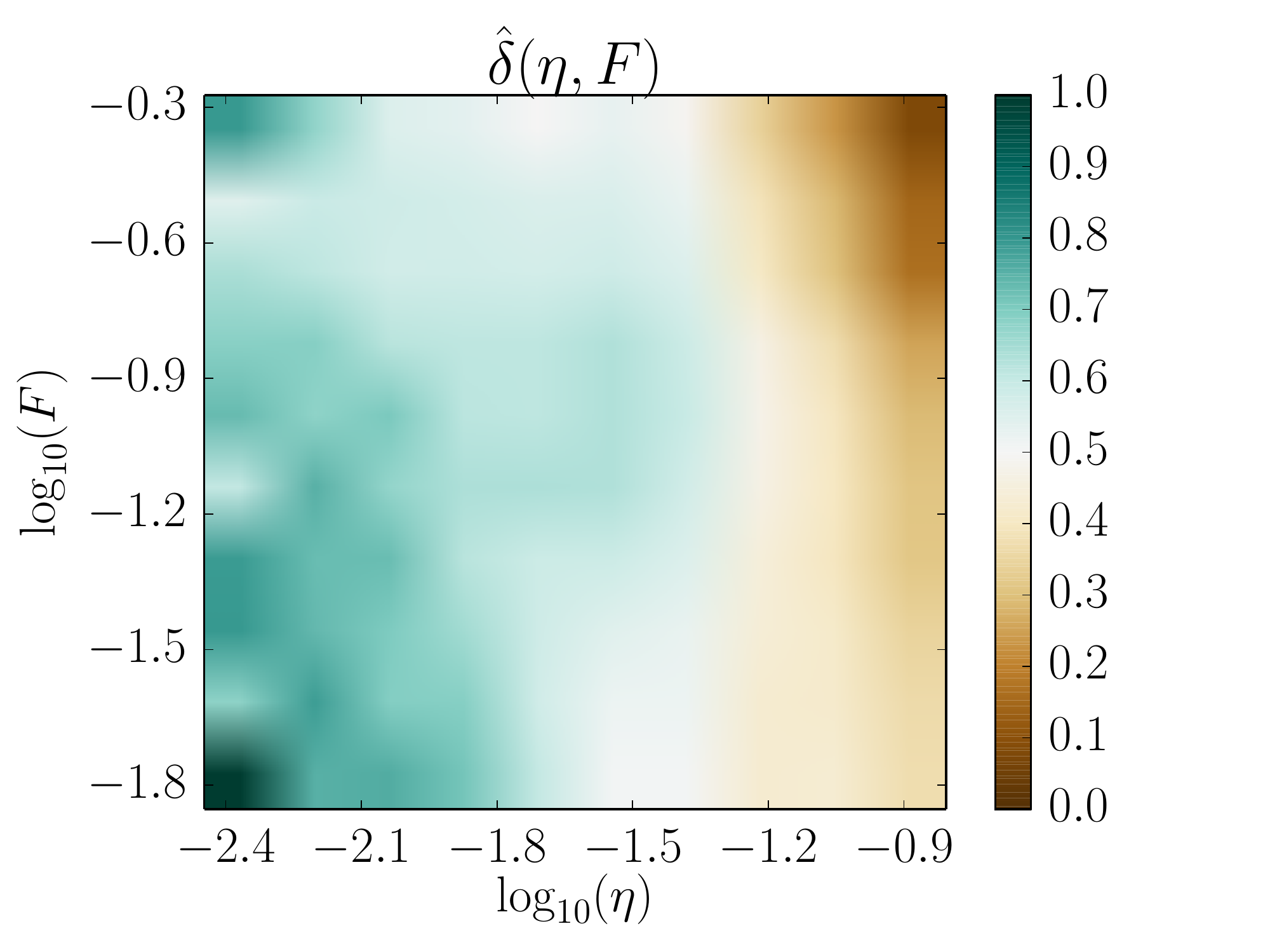}%
 	\includegraphics[width=0.5\textwidth]{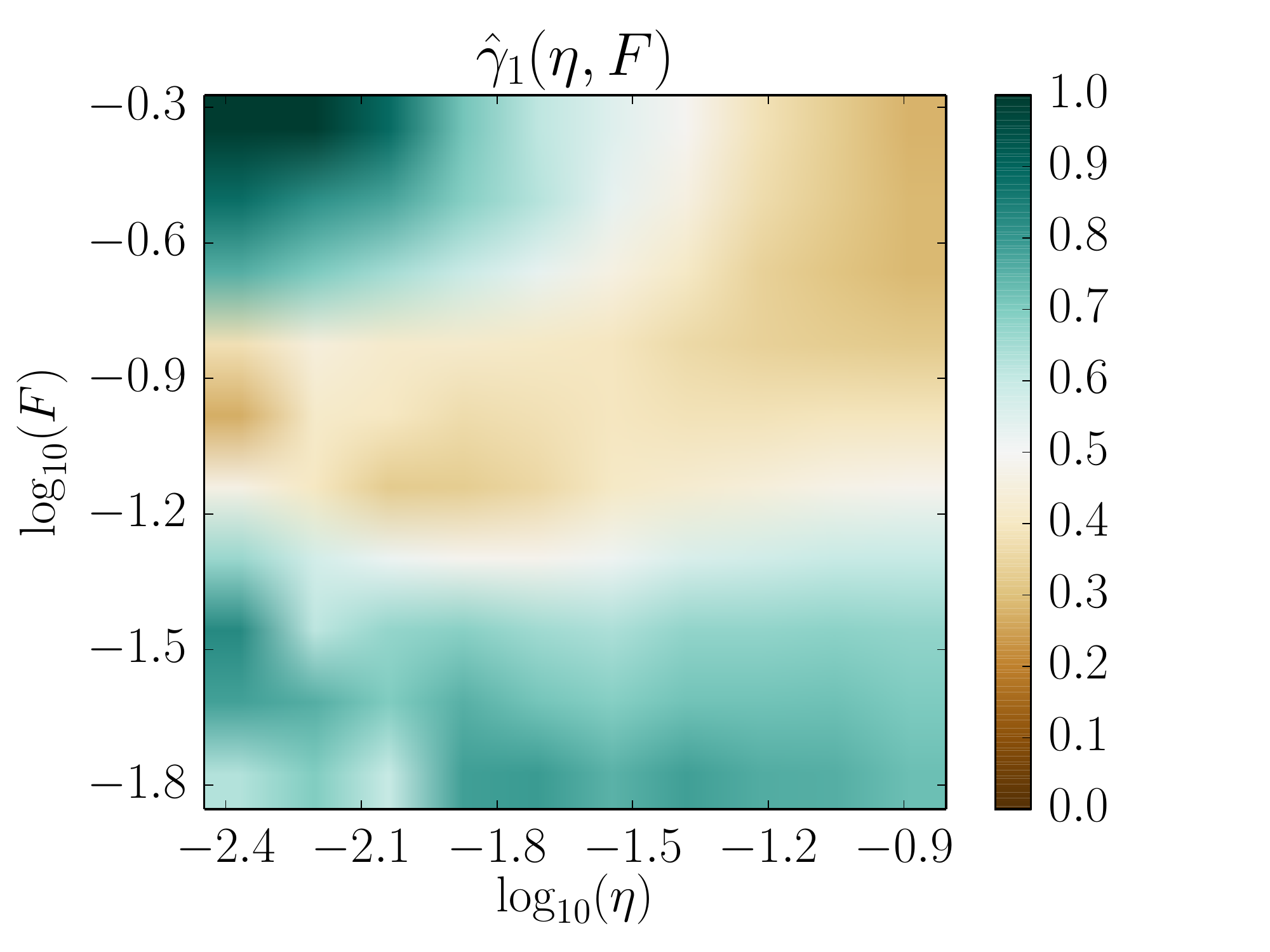}
\caption{Contour plot of the best fitting local exponent $\hat{\delta}(\eta, F)$ (left panel) and $\hat{\gamma_1}(\eta, F)$ measured in the plane $\eta$ -- $F$ adopting the power law function. Green (brown) regions corresponds to locally fitted exponent $\hat{\delta}(\eta, F)$ and  $\hat{\gamma_1}(\eta, F)$ larger (smaller) than 0.5}
   \label{fig_proj}
\end{figure}

To see deviations of the impact surface from a power law from another point of view, we  measure the exponent $\delta$ describing the dependence on $\eta$ and the exponent $\gamma_1$ for the dependence on $F$ via a local non-linear fitting of the functional form $f(\eta, F| Y, \delta, \gamma_1)$ \footnote{
More in detail, we divide the dataset in $n_1=10$ ($n_2=10$) evenly-populated subsets according to $\eta$ ($F$). Each measure in the dataset is labelled by an integer $i_1\in \{1, \dots, n_1\}$ ($i_2 \in \{1, \dots, n_2\}$) according to the subset the measure belongs to. Each measure in the dataset is then labelled by a couple of integers $b=(i_1,i_2)$ identifying a bin. We measure the sample mean of the temporary market impact $I_b^r$ of the metaorders belonging to the same bin $b$ and the average of the relative sample mean of $\eta^r_b$ and $F^r_b$. We define a square identified by the center $c=(c_1,c_2)$, where $c_1 \in \{ 1, \dots, n_1\}$, by considering the bins in the neighbourhood of the center $S_1=\{ c_1-2,c_1-1, c_1, c_1+1, c_1+2\}$ and $S_2=\{ c_2-2,c_2-1, c_2, c_2+1, c_2+2\}$. The square is defined by $S(c_1,c_2)=S_1 \times S_2$.  We select the bins belonging to a square $S(c_1,c_2)$ and we fit the realised temporary market impact $I^r_b$ as a function of the relative realised $\eta^r_b$ and $F^r_b$. We consider a fitting function of the form $f(\eta,F) = C \eta^{\delta}F^{1-\gamma}$. We measure the best fitting parameters $\hat{\delta}(c)$ and $\hat{\gamma}(c)$ as a function of the center of the square $c=(c_1,c_2)$.}.
Figure \ref{fig_proj} shows the local estimation of $\delta$ and $\gamma$ as a function of $F$ and $\eta$.  The structure is very clear.  The green region on the left indicates that the local exponent $\hat{\delta}(\eta,F)$ is consistently larger for small $\eta$, and consistently smaller than $0.5$ for large $\eta$.  The range of variation is significant, with $\hat{\delta}$ varying from roughly $0.1$ to $0.9$.

Similarly, the behavior of the exponent $\hat{\gamma_1}(\eta,F)$ describing the power law scaling on $F$ shows clear structure, though the behavior is a bit more complicated.  For small $F$ the exponent $\hat{\gamma_1}$ is close to one.  For intermediate values of $F$ the exponent is close to $0.3$.  The behavior for larger values of $F$ is more complicated, with high exponents for low participation rates and visa versa for high participation rates.

In conclusion, the non trivial structure appearing in the investigation of the local exponents $\delta$ and $\gamma$ suggests that the logarithmic function $g(\eta, F|a,b,c)$ better describes the temporary market impact surface. The square-root predicted values $\gamma=\delta=0.5$ only works well in the central region of the $\eta-F$ plane.
\\


\subsection{Market impact during the execution of the metaorder}

\begin{figure}[t] 
  \centering
	\includegraphics[width=0.4\textwidth]{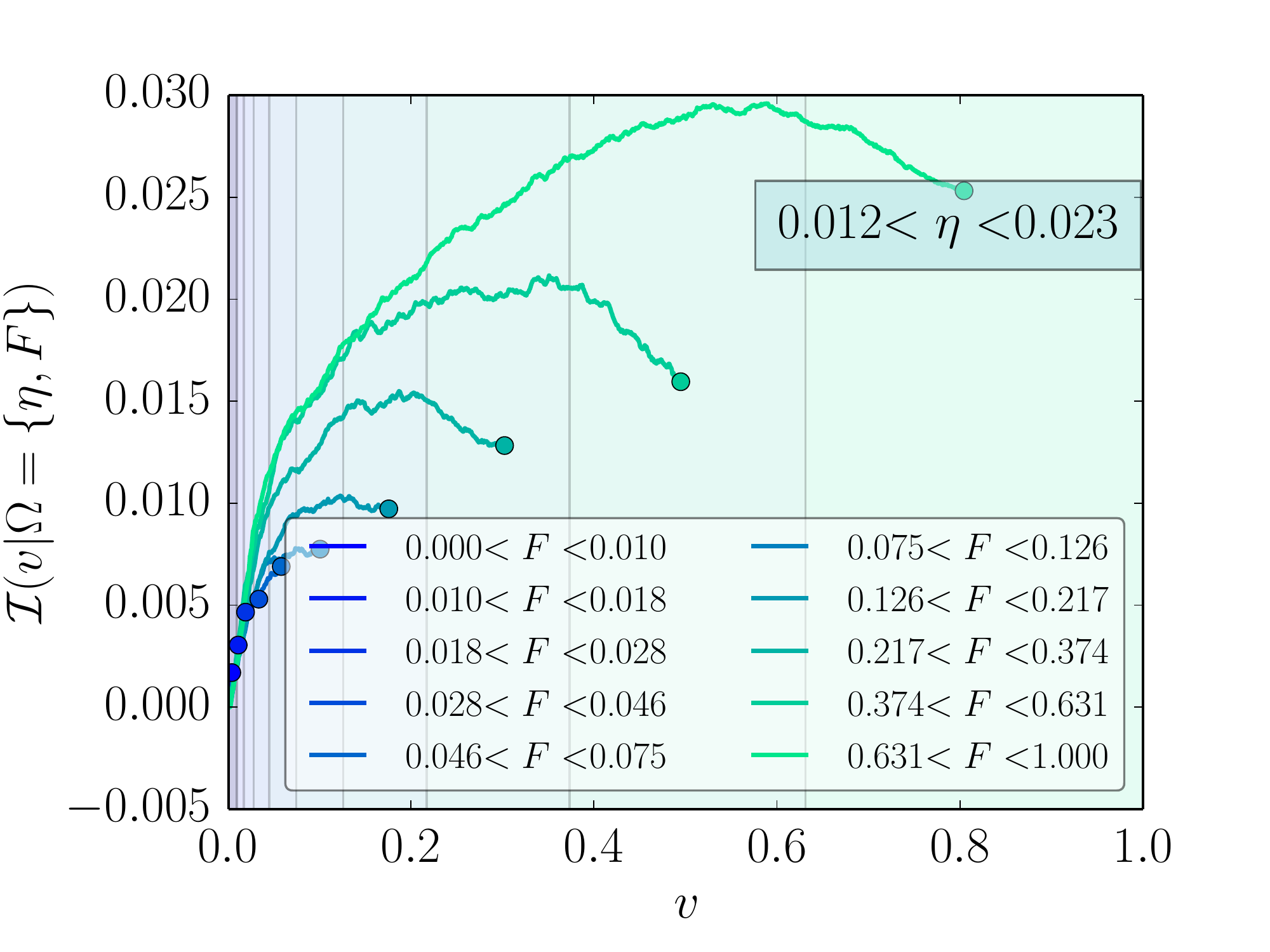}%
	\includegraphics[width=0.4\textwidth]{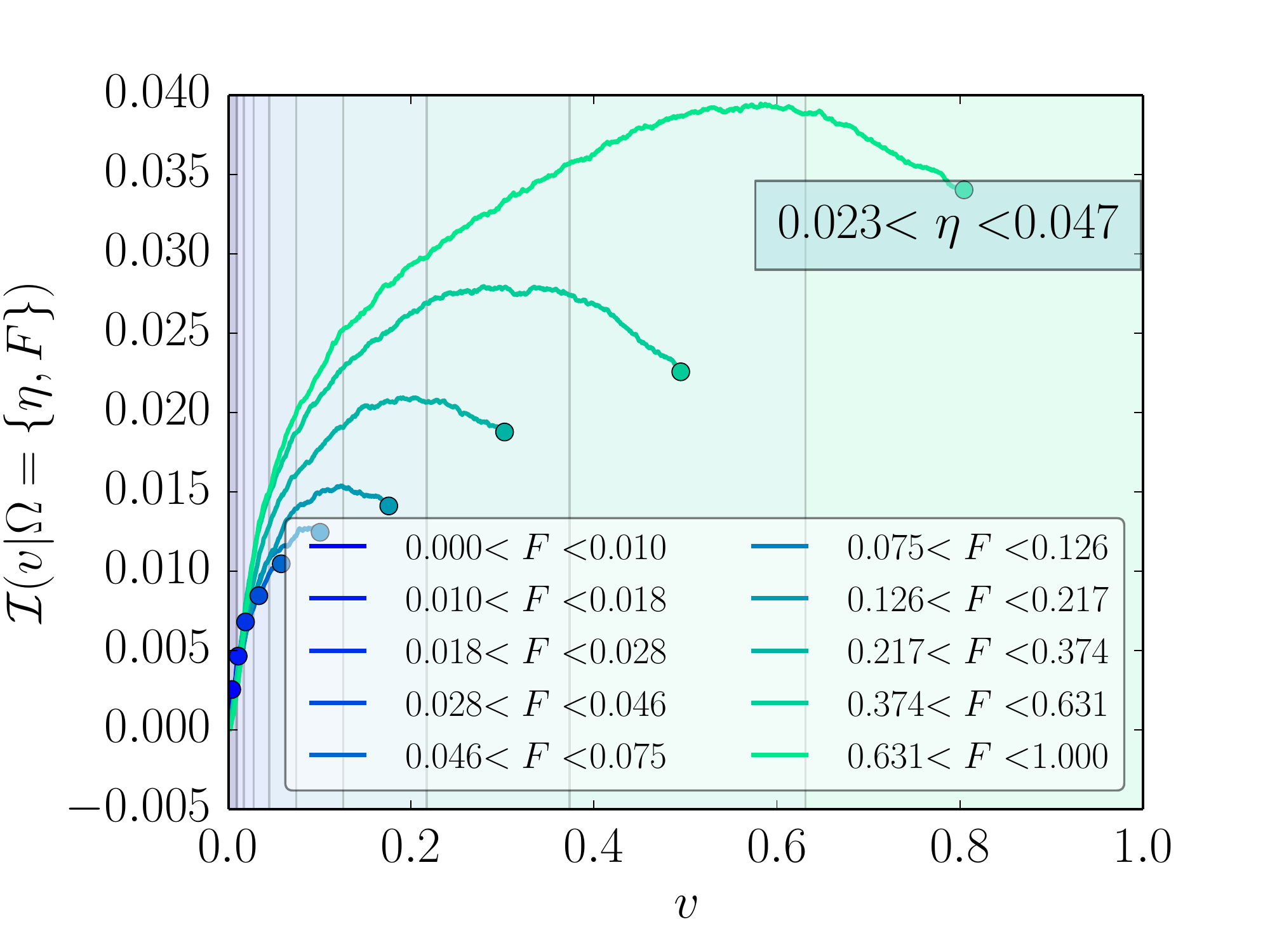}\\
	\includegraphics[width=0.4\textwidth]{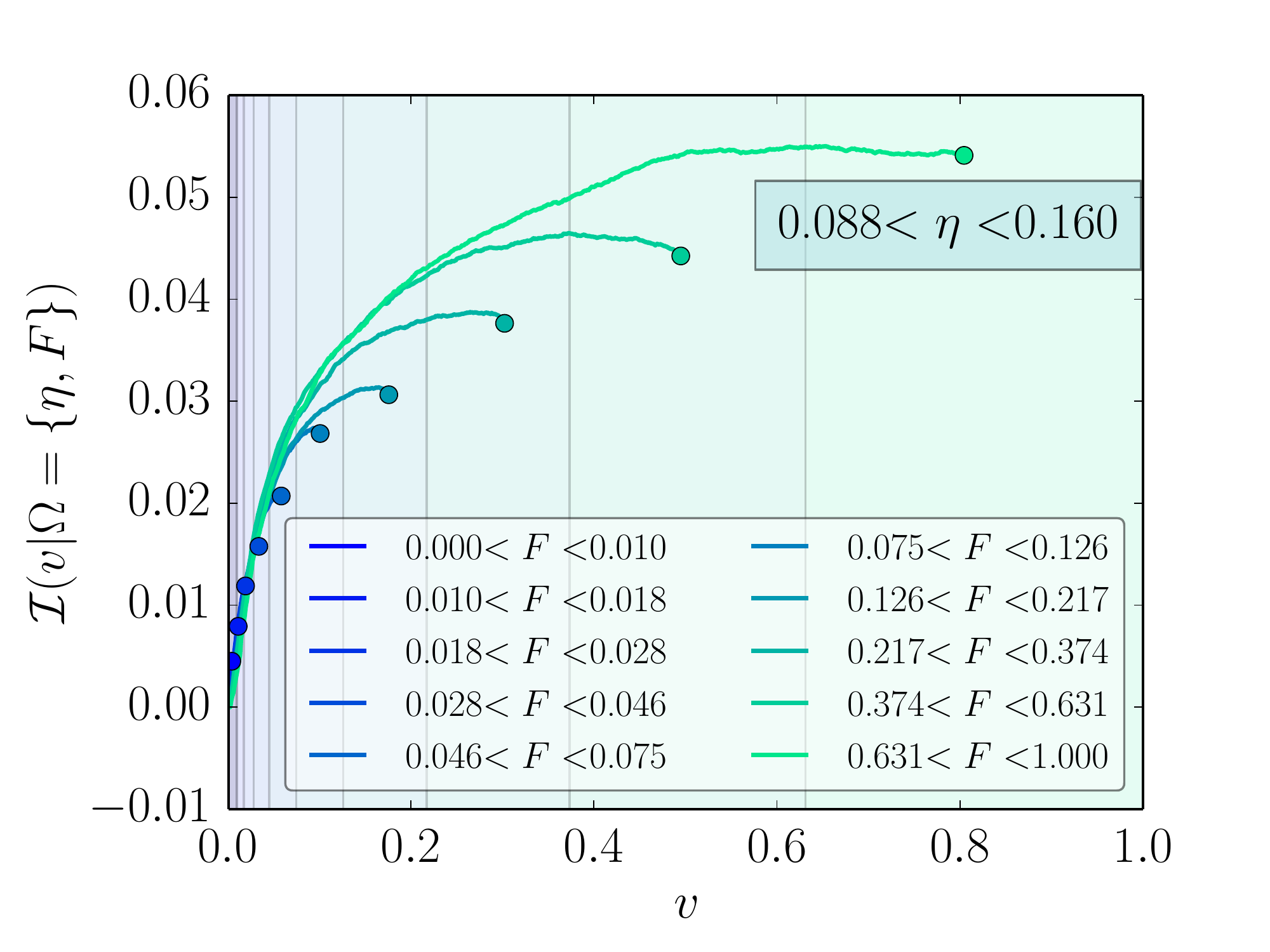}%
	\includegraphics[width=0.4\textwidth]{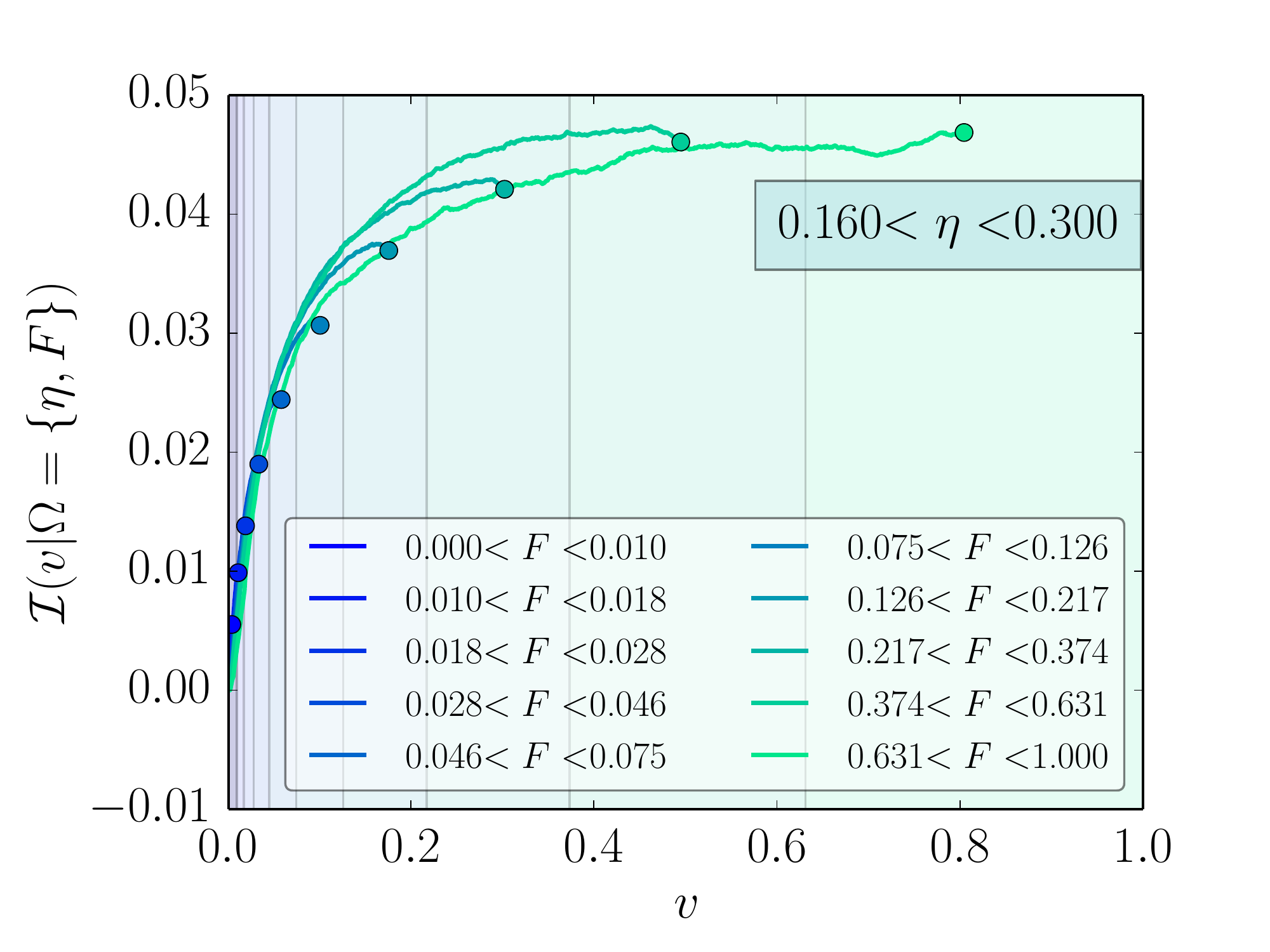}		
  \caption{
  Immediate market impact (solid lines) of metaorders of different participation rate $\eta$ (increasing from the top left to the bottom right panel). Each solid line corresponds to the immediate market impact of metaorders with duration $F_i < F < F_{i+1}$. The temporary market impact is marked by a circle.}
  \label{fig_transient}
\end{figure}

In this section we focus on the immediate market impact, i.e. how the market impact builds up during the execution of a metaorder.  We consider the following question: Given two metaorders with the same participation rate $\eta$ and different durations $F_1$ and $F_2$ ($F_1<F_2$), should we expect that the market impact reached at time $F_1$ is the same for the two metaorders? A priori, when the first is at its very end while the second one is still being executed, they should be indistinguishable from the point of view of market impact, since the public information available up to this time is the same for both metaorders (see also the discussion in the very recent paper of \cite{Iuga14}).


In Figure \ref{fig_transient} we show the result of our analysis\footnote{
A methodological comment is in order. The Ancerno database does not provide the number, time, volume, or price of each individual transaction (or child-order) through which the metaorder has been executed. For this reason we follow the price dynamics by using the public price information with one-minute time resolution.}.
The four panels refer to four different bins of the participation rate $\eta$. 
In each panel we consider 10 bins of duration $F_i$. For each of them a line represents the average price path during the execution $\mathcal{I}(v | \Omega=\{ \eta, F\})$ for metaorders with duration $F_i < F < F_{i+1}$. 
The circles are the temporary impacts for metaorders of volume time duration $F_i$. The figure clearly gives a {\it negative} answer to our question. 
In fact, the price trajectories deviate from the temporary market impact described by the circles. For small participation rates this effect is more evident and price trajectories are well above the immediate impact. Notice also that in some cases the price reverts {\it before} the end of the metaorder\footnote{Note that, with this method, we are able to investigate in full detail the the market impact path in the early stage of the execution, because for orders with duration $F_i < F < F_{i+1}$ we follow the price up to $v=F_i$, missing the very last part. As a consequence, in figure \ref{fig_transient} the magnitude of the reversion of the market impact path during the execution is underestimated. On the contrary, this feature becomes more evident in the analysis performed in the following section, see Figure \ref{fig_permanent} and \ref{fig_cnt_2}. In that case, we follow the market impact path with great attention to the late stage of the execution. We will observe that long metaorders with small participation rate present a strong reversion, as clearly visible in Figure \ref{fig_cnt_2}.\label{foot}}.  A similar behavior  was very recently shown in Ref. \cite{Iuga14}. For larger participation rates the price trajectories become closer and closer to the circles representing the temporary impact. 

The discussion presented in Section \ref{sec_models} helps us understanding this behaviour. We have seen that within the Almgren-Chriss model, the market impact trajectories deviate from the temporary market impact surface if the execution profile deviates from the VWAP trading profile. Front-loaded execution profiles, used for example by risk averse investors, generate market impact trajectories that stay above the market impact surface as shown in figure \ref{fig_ac}. However the Almgren-Chriss model is not able to reproduce some of the main features of market impact since it predicts a linear market impact. The propagator model, on the contrary, better reproduces the concavity of the market impact surface and consistently makes it possible to recover the square-root law describing the market impact curve. As we have seen in section \ref{sec_prop}, the model predicts that, also in this case, front-loaded execution schemes have market impact trajectories that depart from the impact surface and reach it from above. Consistently, the presence of a decaying kernel for the impact, makes it possible to reproduce the fact that price starts to revert {\it before} the end of the execution. If the trading pressure is softer than the market recovering force, market impact starts reverting. 

These results strongly suggest that the typical trading profile used by the brokers in our database is not the widespread VWAP  trading profile but a front loaded execution scheme. This might be due to risk aversion or in order to avoid losing a profit opportunity. In fact, if the price is expected to increase, it is better to buy more at the beginning of the metaorder and less at the end. Unfortunately, these alternatives can not be tested within the information contained in the Ancerno database. 
\\

\subsection{Impact decay and permanent impact}

In this section we consider the temporal dependence of the price after the execution of the metaorder, i.e. how the market impact relaxes. The long term limit of the price, when all the temporary effects have dissipated, is called permanent impact. Recently there has been a debate about the value of the permanent impact and the dynamics of the price after the end of the metaorder.  Under the assumption that  the metaorder size distribution has a power law tail with exponent $1.5$, the model of Farmer et al. \cite{farmer2013efficiency} predicts a decay of the impact to a permanent value of roughly $2/3$ of the peak impact.  Note however, that this assumption does {\it not} appear to be met in our data -- as seen in Figure \ref{fig_stat}, $\pi$ clearly does not have a power law distribution.  This has serious consequences, since in this case their theory does not predict a square root market impact function.  See Section \ref{fundamentalModels} for a more complete discussion.


Here we consider the market impact trajectory $\mathcal{I}(v|\Omega = \{ \eta,F\})$ after the end of the execution of the metaorder, i.e. for $v>F$. In order to compare metaorders with different durations $F$, we rescale time as $z=v/F$. In this way at metaorder completion it is $z=1$ independently from the metaorder duration $F$.  We also rescale the market impact trajectory by dividing by the market impact at the end of the execution of the metaorders $\mathcal{I}(v=F|\Omega = \{ \eta,F\})$, i.e.
\begin{equation}
\mathcal{I}_{ren}(z|\Omega = \{ \eta,F\}) := \frac{\mathcal{I}(z|\Omega = \{ \eta,F\})}{\mathcal{I}(z=1|\Omega = \{ \eta,F\})}.
\end{equation}
The decay of the impact is presumably less dependent on the execution scheme than the immediate impact, and therefore its study could be used, at least in principle, to investigate how well the propagator model describes the price dynamics. As we will see, this is strictly true only if we can neglect the order flow of other metaorders (for example if the participation rate of the conditioning metaorder is large enough).

\begin{figure}[t]
  \centering
	\includegraphics[width=0.6\textwidth]{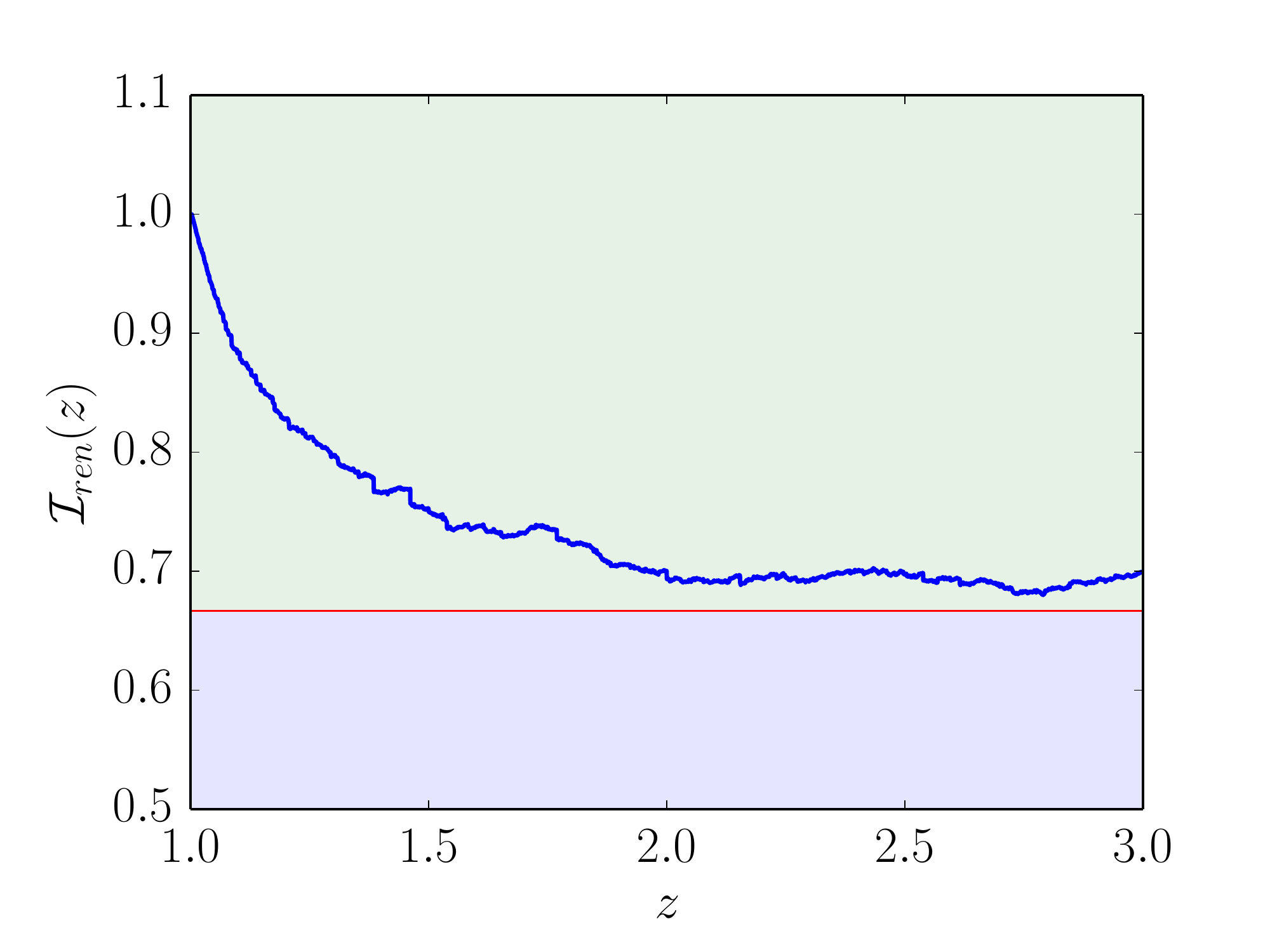}
\caption{Decay of temporary market impact after the execution of the meteorder. We follow the normalised market impact path $\mathcal{I}_{ren}(z)$ as a function of the rescaled variable $z=v/F$, without conditioning on any variable. The market impact path of each metaorder is followed also in the following day. The red horizontal line corresponds to 2/3, as predicted by the model of Farmer et al. \cite{farmer2013efficiency}.}
\label{fig_permanent_all}
\end{figure}

We first consider all the metaorders together and we compute the average rescaled path followed by the price after the end of the metaorder. The result is shown in Figure \ref{fig_permanent_all}. We observe that the price decays toward a value which is remarkably close to (though slightly higher than) $2/3$ of the peak impact. This is agreement with the results obtained for example in \cite{moro2009market,bershova2013non}. 

\begin{figure}[t]
  \centering
	\includegraphics[width=0.38\textwidth]{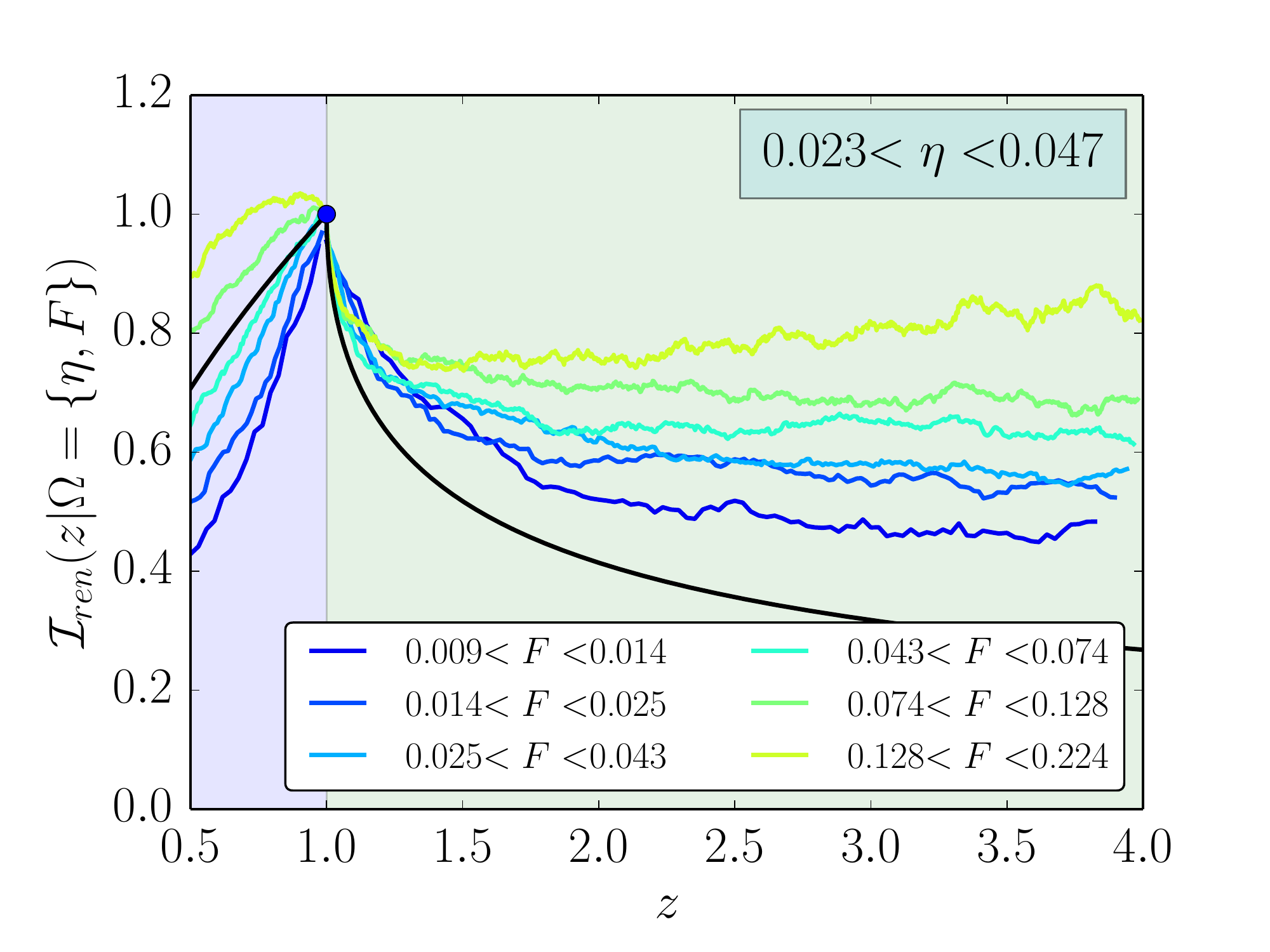}%
	\includegraphics[width=0.38\textwidth]{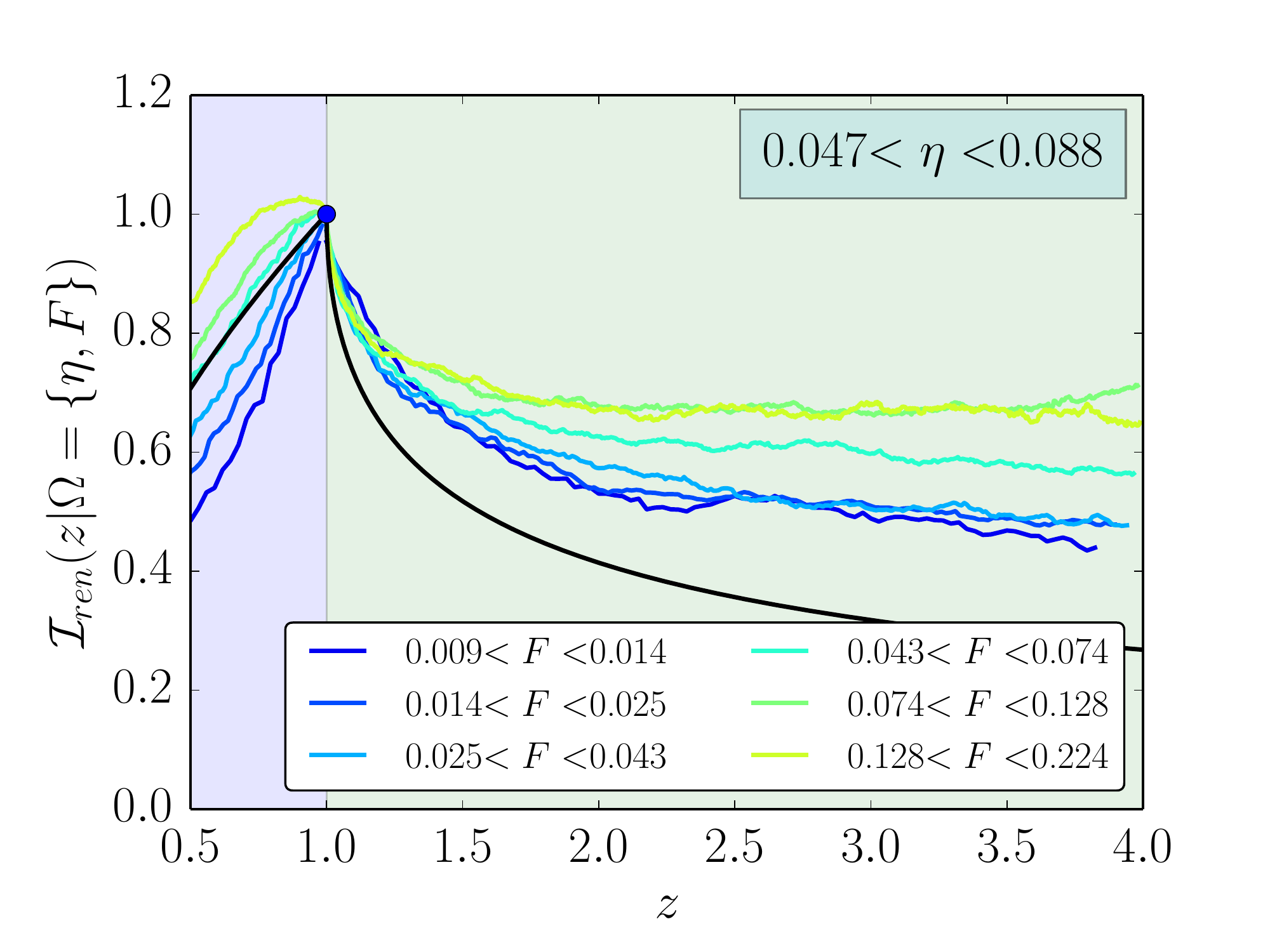}\\
	\includegraphics[width=0.38\textwidth]{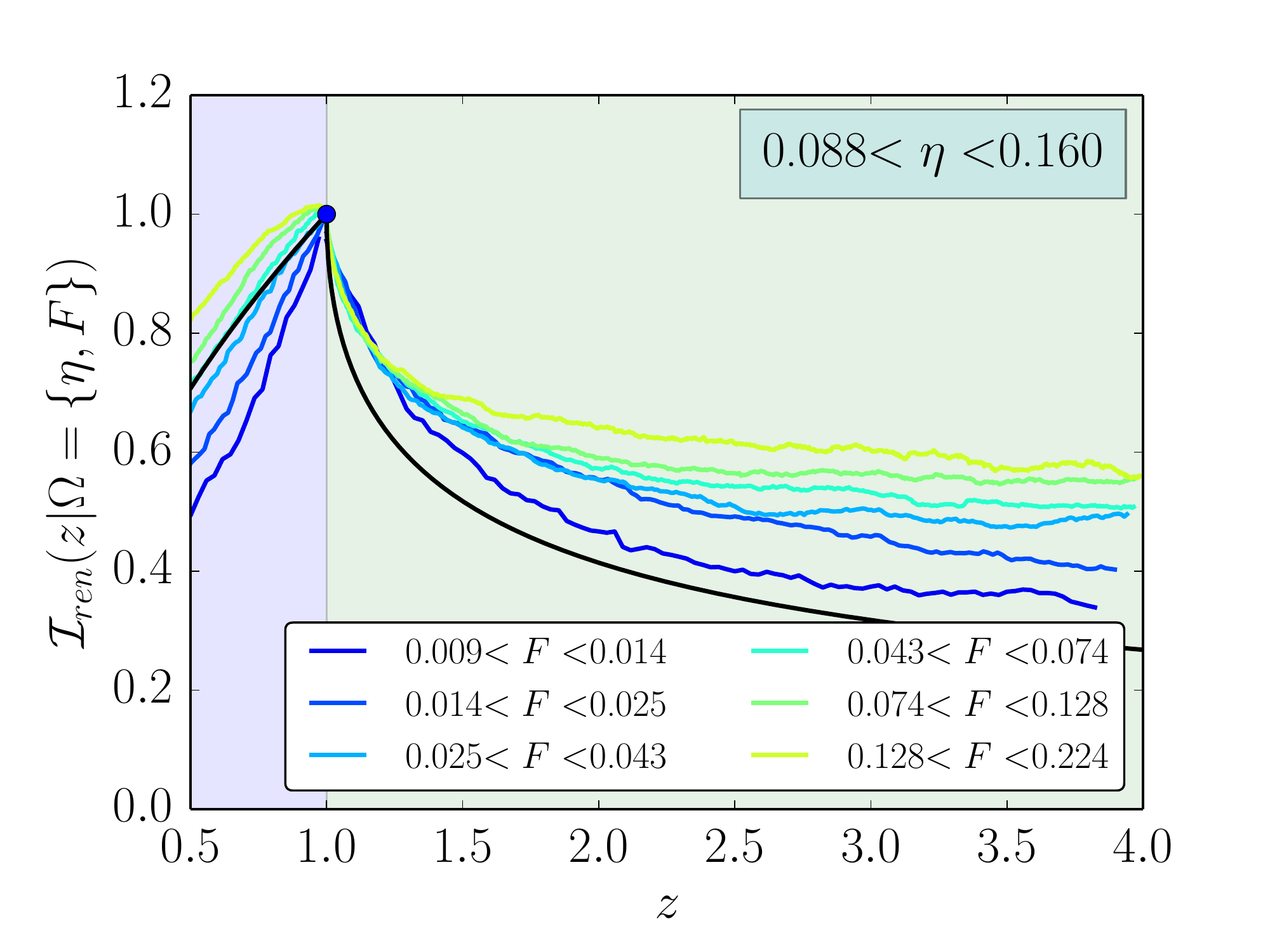}%
	\includegraphics[width=0.38\textwidth]{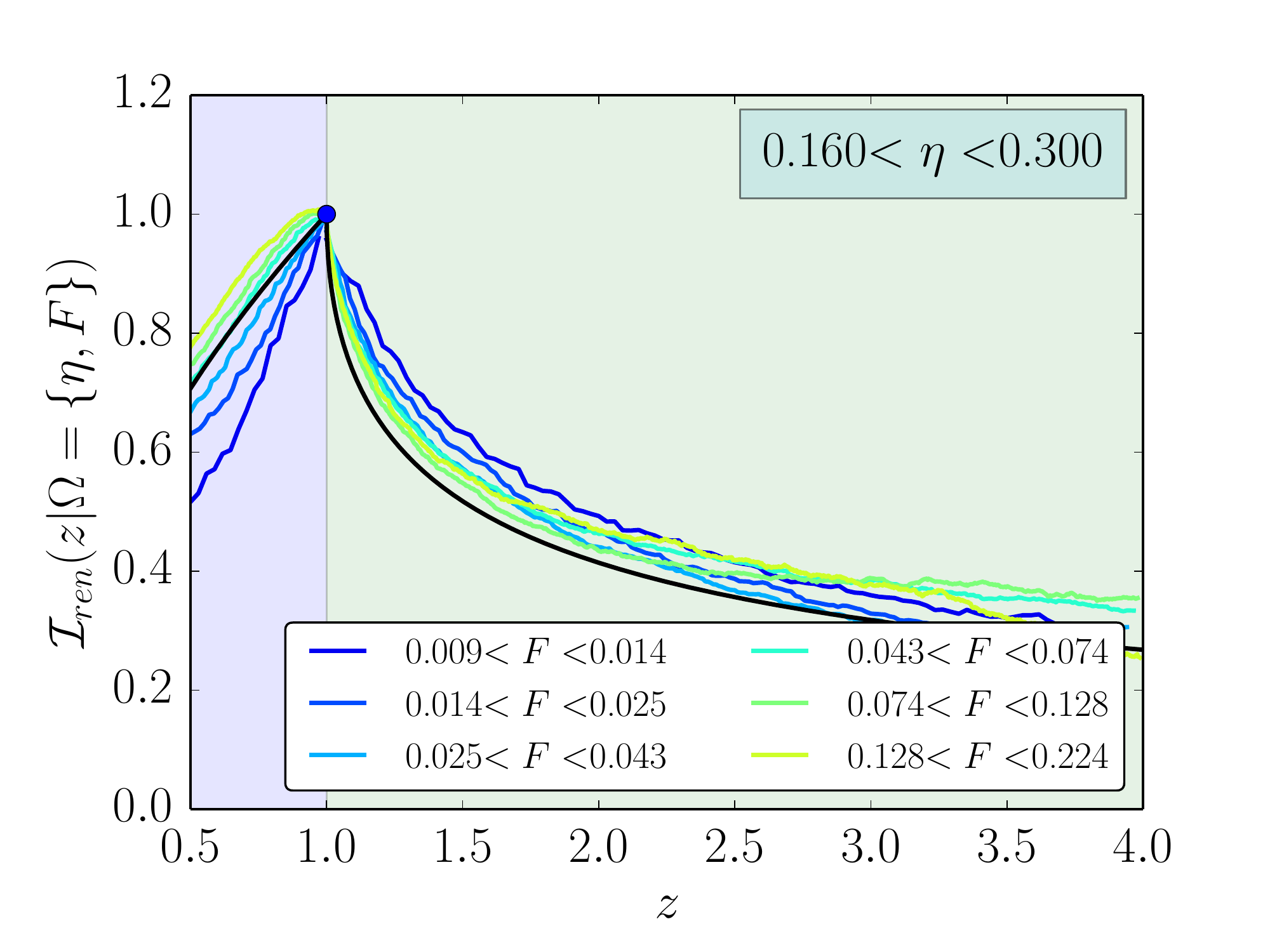} 
	  \caption{Decay of temporary market impact after the execution of a metaorder. We follow the normalised market impact path $\mathcal{I}_{ren}(z|\Omega = \{ \eta,F\})$ as a function of the rescaled variable $z=v/F$.  Within each panel the solid lines correspond to the average market impact trajectory for metaorders with different durations $F$; the four panels correspond to different participation rates $\eta$.   We consider the price dynamics up to the end of day when the metaorder was placed. The black line corresponds to the prediction of the propagator model with $\delta = 0.5$.  Overnight returns and the price path of subsequent days are not considered in our analysis. }
\label{fig_permanent}
\end{figure}

The large size of our metaorder database allows us to perform an analysis of the price decay conditioning on the duration and participation rate. Figure \ref{fig_permanent} shows the results. The four panels refer to increasing values of the participation rate. In each panel the market impact path of metaorders with several durations is presented. We follow the relaxation of the market impact trajectories up to three times the duration of the metaorder, but we avoid introducing overnight returns and following the price on subsequent days. In each panel we also show the prediction of the propagator model for the market impact trajectory when $\gamma = 0.5$ (black line). 

The figure shows that the price decay and its long term limit depend on $\eta$ and $F$. For small participation rates (top panels) the average permanent impact (across durations) is close to $2/3$ peak impact. However this is also the regime where we observe the strongest dependence of the permanent impact on $F$.  Longer metaorders relax more slowly than shorter metaorders, and at the end of the period examined remain at higher price levels.   This effect is bigger for smaller participation rates\footnote{\
This observation that metaorders with low participation rate revert more slowly is consistent with the notion that reversion depends on detection by others of the presence or absence of the order.  The beginning or end of a low participation rate order is more difficult to detect, and should require more time for a given level of certainty, giving a more sluggish reaction to completion of the order.  Of course there may also be other explanations.}.
On the contrary, for the largest participation rates the renormalised market impact paths of metaorders are all very similar. The market impact relaxes toward zero and we do not observe any flattening of the curve in the considered time window.  Quite interestingly, in this regime the market impact decay is well described by the prediction of the propagator model with $\gamma=0.5$, while for small and intermediate participation rates the price is systematically higher than the value predicted by the propagator model.

\begin{figure}[tc] 
  \centering
	\includegraphics[width=0.38\textwidth]{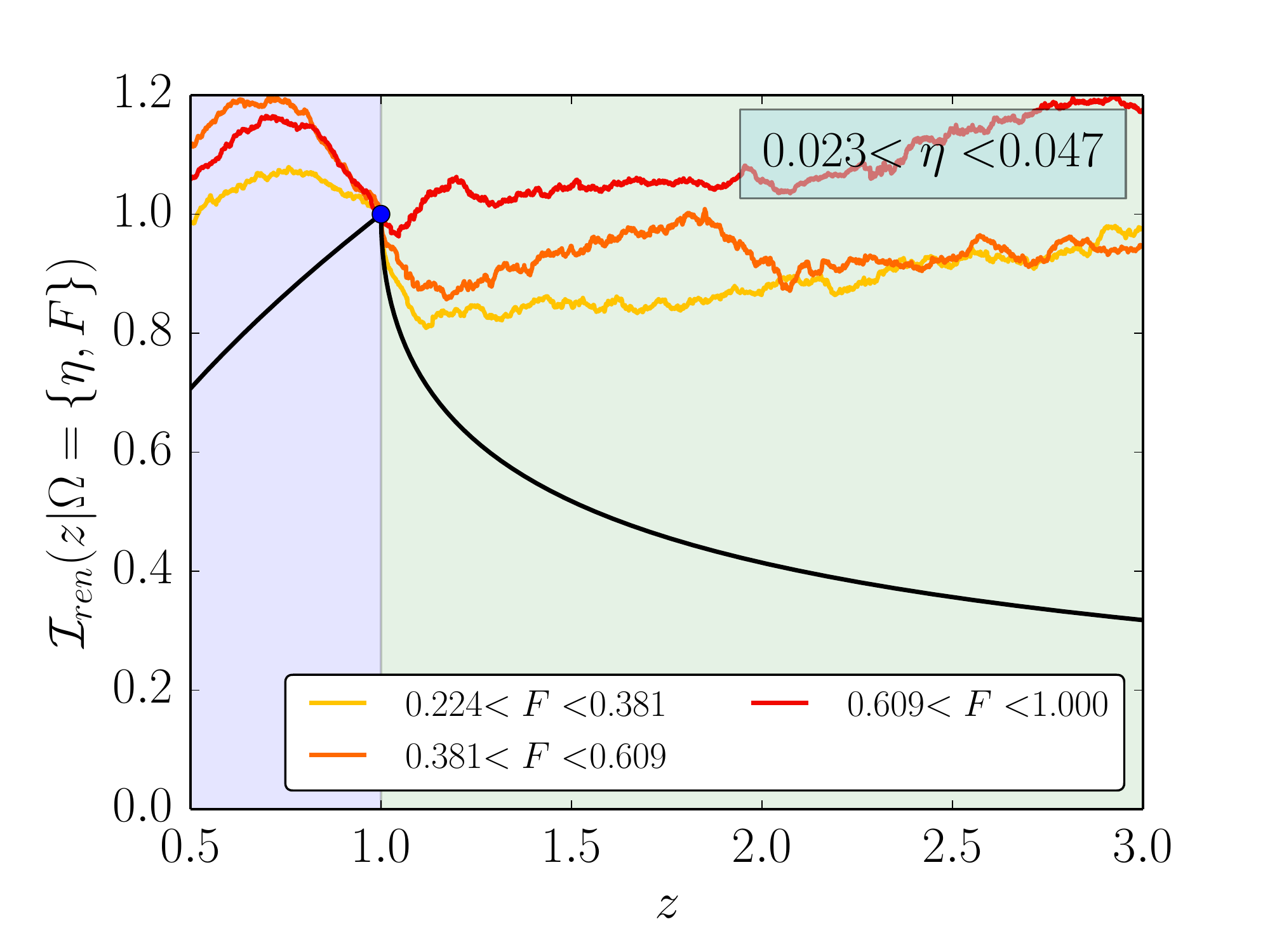}%
	\includegraphics[width=0.38\textwidth]{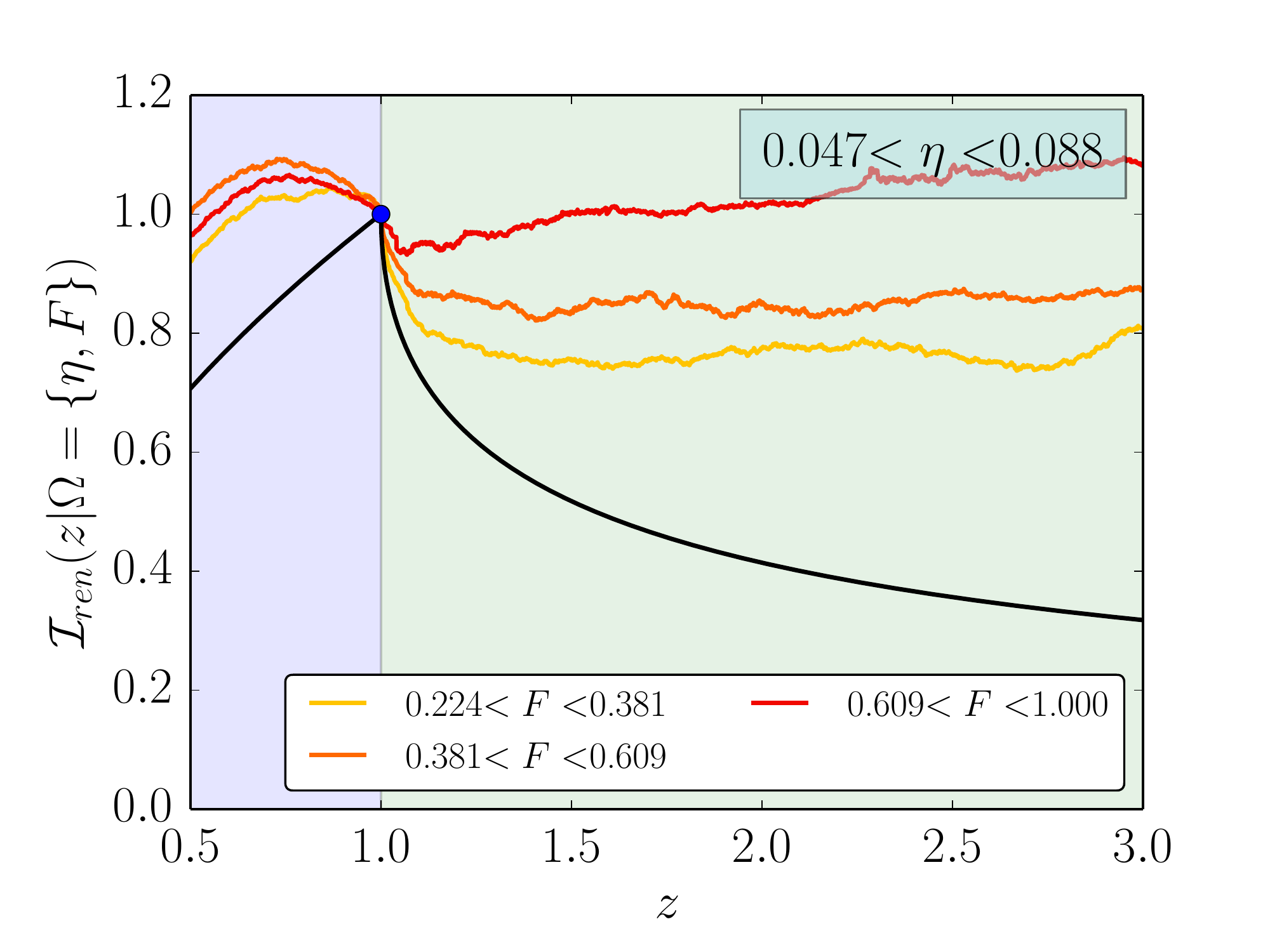}\\
	\includegraphics[width=0.38\textwidth]{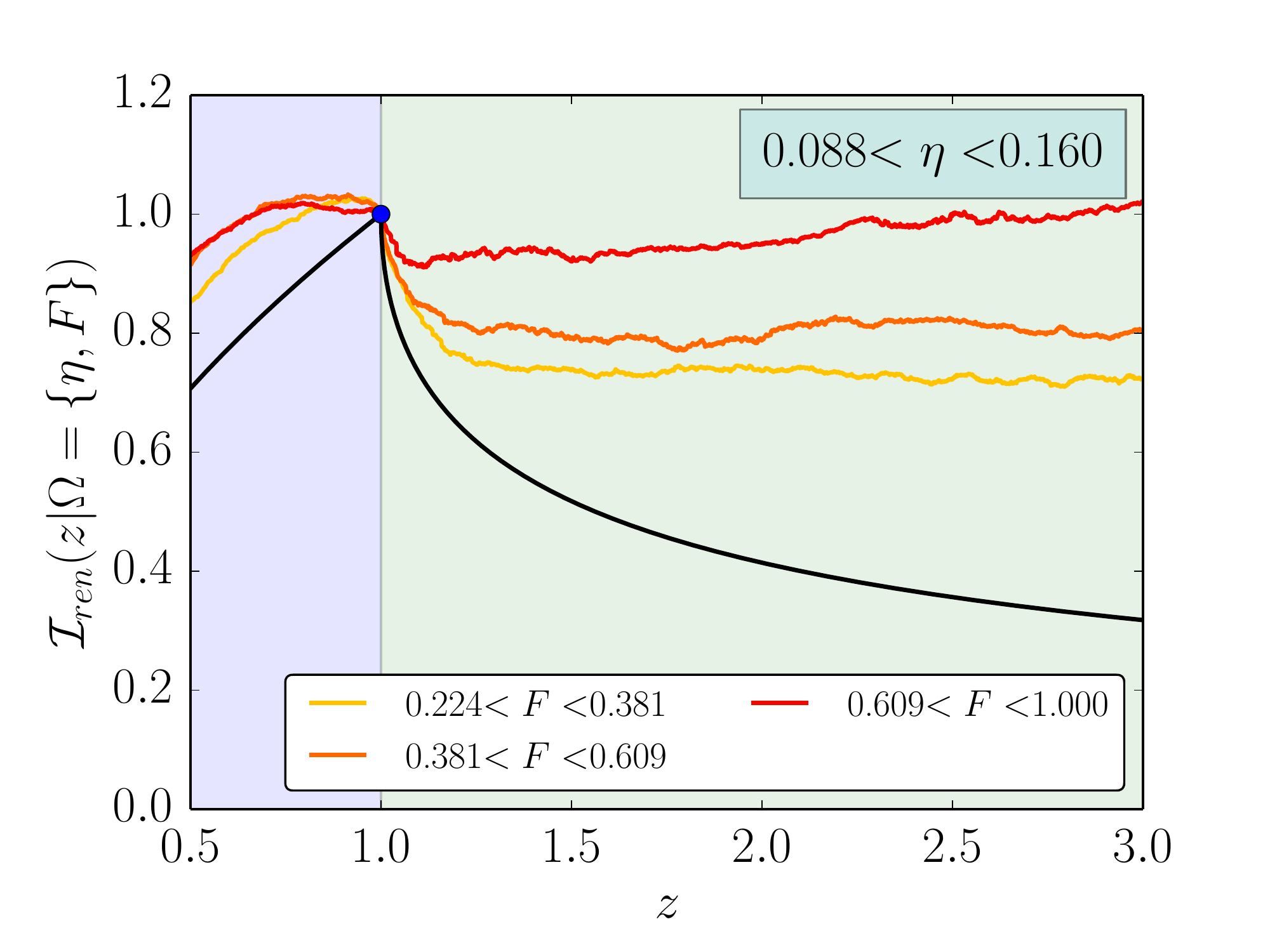}%
	\includegraphics[width=0.38\textwidth]{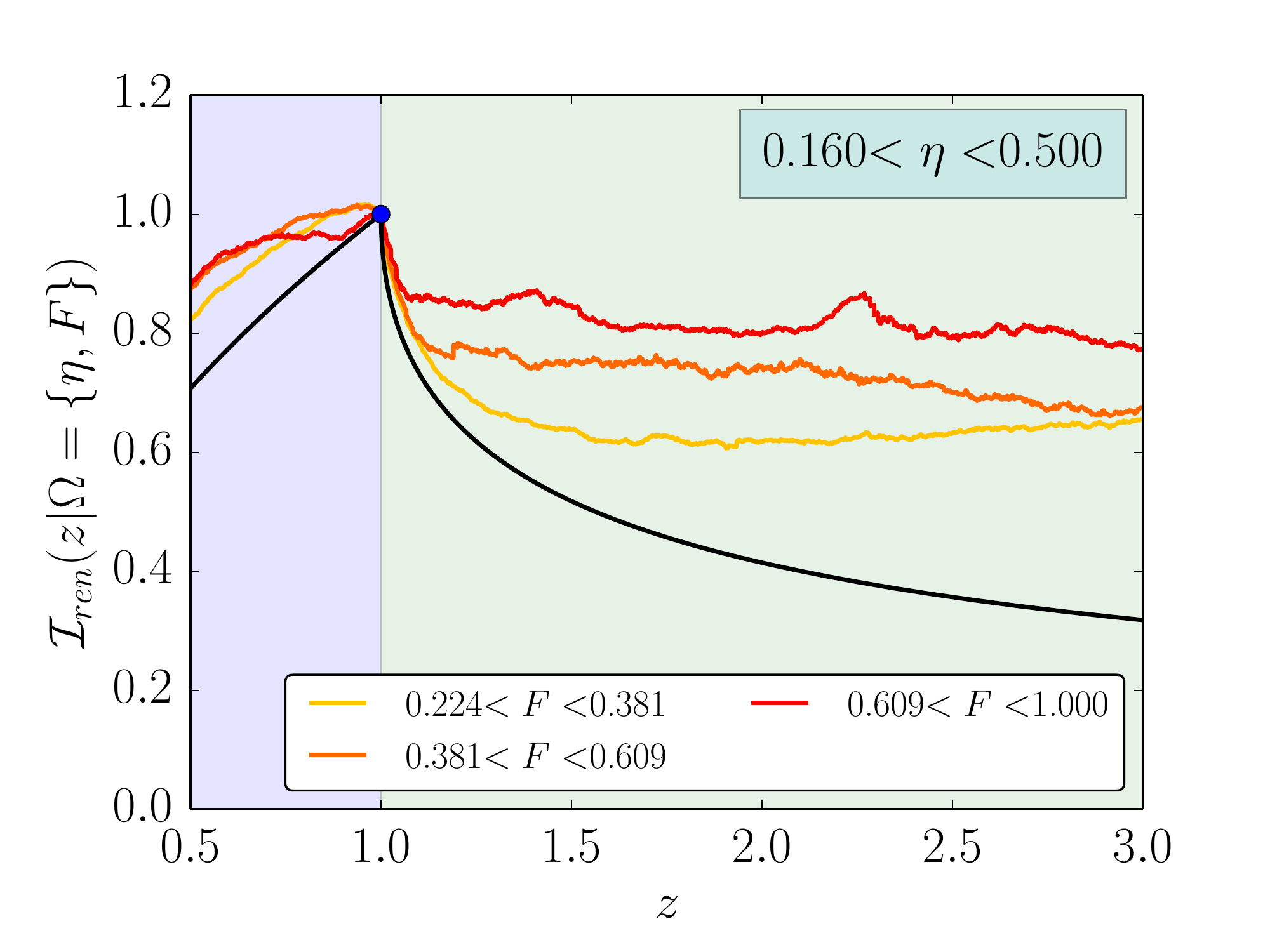} 
 \caption{Decay of temporary market impact after the execution of the metaorder.  We follow the renormalised market impact path $\mathcal{I}_{ren}(z|\Omega = \{ \eta,F\})$ as a function of the rescaled variable $z=v/F$. Each solid line corresponds to the average market impact trajectory computed on metaorders characterised of low (top row) and high (bottom row) participation rate $\eta$ and several durations $F$ (see legend). We consider metaorders with long durations $0.224<F<1$ and we follow the price path also in the following days (in contrast with the analysis of figure \ref{fig_transient}).  }
  \label{fig_cnt_2}
\end{figure}

We have also performed the previous analysis following the market impact decay on subsequent days. This allows us to include metaorders with longer duration in the analysis. 
Considering metaorders with the same duration as before, we observe that the global picture changes slightly only for metaorders with very large participation rate. In the other cases it is approximatively unchanged (data not shown). Considering metaorders with even longer duration, see Figure \ref{fig_cnt_2}, we observe the appearance of clear plateaux with height 0.8 - 1 times the peak impact. It is worth also noticing that here, but also in the top left panel of Figure \ref{fig_permanent}, the reversion of the price before the end of the metaorder is much more clearly visible, as explained in footnote \ref{foot}.
Note that the market impact trajectories of metaorders in this analysis often contain the overnight return (contrary to the previous analysis). As seen by \cite{brokmann2014}, we observe that the price decay essentially stops when the trading day ends. However the presence of overnight returns increases significantly the already large noise in the determination of permanent market impact.

\subsubsection{The role of metaorder autocorrelations}

The picture emerging from the previous analysis can be partly clarified by taking into account the autocorrelation of the sign of metaorders.  Positive autocorrelations in the signs of metaorders will make the market impact of a single metaorder relaxation artificially high, as it becomes impossible to isolate metaorders from each other.  Moreover this effect is stronger for longer metaorders, since the probability of overlapping with other metaorders is larger.  It is also larger for lower participation rates, since the market impact is easily overwhelmed by that of metaorders with larger participation rates. On the contrary, we expect that the effect is milder for shorter metaorders, because of the lower probability of overlap, and larger participation rates, because the effect of metaorders with lower participation rates on price becomes negligible.  

\begin{table}[ht]
\begin{center}
\begin{tabular}{| r | r | r | r | r |}
\hline
duration (mins) & number & \# overlaps & same sign & opposite sign  \\
\hline                       
0 - 10  & 368,484  & 1.7  & 0.548  & 0.452 \\
10 - 25  & 117,756  & 3.0  & 0.553  & 0.447 \\
25 - 50  &  71,031  & 4.6 & 0.553  & 0.447 \\
50 - 100  & 52,931  & 6.6  & 0.547  & 0.453 \\
100 - 200 &  43,884  & 8.0 &  0.546 & 0.454 \\
200 - 390 & 49,411  & 9.3 & 0.543 & 0.457 \\
\hline
0 - 390 & 703,497 & 3.54 & 0.548 & 0.452 \\
\hline
\end{tabular}
\end{center}
\caption{{\it An analysis of overlap of metaorders.}  We consider the metaorders with participation rate $\eta > 0.005$ traded on the 100 most populated stocks from January 2007 to December 2009. This set has 703,497 metaorders. We consider nonintersecting bins according to the duration of the metaorders (first column) and their relative number (second column).  For each metaorder we consider the time interval from the beginning up to $3$ times its duration. We count the metaorders in the whole set overlapping with the selected time interval. For each subset we report the average number of overlapping metaorders ( \# overlaps, third column). As expected, the number of overlaps increases with the duration of the metaorder. We then measure the fraction of the overlapping metaorders which have the same or opposite sign as the selected metaorder (fourth and fifth columns). We observe that a constant average fraction ($\sim 55 \%$) of the overlapping metaorders have the same sign, independent of the duration. This finding quantifies the autocorrelation of the sign in the time series of the metaorders. }
\label{tab_anom}
\end{table}

The overlap of the metaorders present in our database is summarized in Table \ref{tab_anom}. We observe that, considering the time interval from the beginning of the metaorder up to three times its duration after the end of the execution, on average, a given metaorder overlaps on average with 3.5 other metaorders. As expected, the average number of overlapping metaorders is larger for longer metaorders (around 2 for the shortest ones and around 10 for the longest ones). On average, $55 \%$ of these metoarders have the same sign. This implies that, on average, a metaorder is surrounded by more metaorders of the same sign than of the opposite signs, and this effect enhances the measured impact.

Very recently Ref. \cite{brokmann2014} considers trades from the same fund and traded following a signal and show that they present a strong autocorrelation in time. The authors suggest that a positive autocorrelation of sign of the metaorders can keep the impact artificially high. They suggested a method to deconvolve their own trades to remove both their own impact and the information, finding zero permanent impact on the time scale of 15 days. It is important to highlight that, although consistent with \cite{brokmann2014}, our measure of the autocorrelation of metaorders is obtained by using an extensive database covering the trading activity of many different investors, rather than all the metaorders of the same fund. Thus our analysis points out a herding among funds in their trading of metaorders, rather than metaorders by the same institution in the attempt to exploit medium term signals as in Ref. \cite{brokmann2014}.


The positive autocorrelation of metaorder signs qualitatively explains the findings on price decay. Market impact trajectories of metaorders with very large participation rate are negligibly perturbed by the other metaorders and their trajectories are roughly independent of duration (bottom right panel of Figure \ref{fig_permanent}).
Moreover, the market impact trajectory is quite well described by the propagator model. On the other hand, we have seen that the market impact trajectory of metaorders with lower participation rate and longer durations deviates from each other and from the prediction of the propagator model (top panels of Figure \ref{fig_permanent}). We speculate that, in this case, the market impact trajectories are kept artificially high by the effect of other metaorders with the same sign and non-negligible participation rate. The fact that this effect is stronger for low participation rates is consistent with our explanation. Although interesting and worthy of investigation, a more in depth analysis of this aspect is beyond the scope of this paper.

\section{Implications for fundamental models \label{fundamentalModels}}

One of the  motivations for this paper is to test fundamental theories for market impact.  In this section we review these theories and discuss their possible implications in relation to the results presented here.  We also offer some caveats, discussing possible effects that might distort our results. 

\subsection{The latent order book approach of Toth et al.}  

Toth et al. \cite{toth2011anomalous} present a theory for market impact based on the concept of a latent order book.  The key idea is that the true order book does not reflect the actual supply and demand that are present in the market, due to the fact that participants do not reveal their true intensions.  They show that for prices to be diffusive, i.e. for the variance to grow linearly with time, it is necessary for the latent order book to have a linear profile around the current price, which implies a square root impact function. This is supported by simulations of a simple agent-based model.   They make no prediction about how prices should relax after execution is completed, though in subsequent empirical work this group suggests that once the predictive advantage of a trading strategy has been removed the price relaxes slowly to zero \cite{brokmann2014}.

That fact that we observe a logarithm for temporary impact appears to contradict the theory of Toth et al.  While we do observe that the square root is an approximation over part of the range, we see substantial deviations.  In addition the fact that we observe an impact surface with logarithmic dependence on $\eta$ and $F$ separately is not consistent with their theory.   However see the caveats given below, as well as the discussion of the implications for the latent order book in Section~\ref{virtualOrderBook}.

\subsection{The fair pricing approach of Farmer et al.}

Farmer et al \cite{farmer2013efficiency} derive a fair pricing principle that, when combined with the martingale property of prices, predicts that the average execution price should equal the final price when the metaorder has completed and prices have been allowed to relax.  This is done by deriving a Nash equilibrium between informed traders and liquidity providers, in a setup that requires much stronger assumptions than the theory of Toth et al. above \cite{toth2011anomalous}.  (This model can be viewed as an extension of Kyle's original model, but with more realistic assumptions.  Farmer et al. assume batch executions and that the beginning and end of metaorders is known by market participants.  The functional form of market impact depends on the distribution of metaorder sizes.  Under the assumption that the cumulative distribution of metaorder sizes is a power law with exponent $-3/2$ they predict a square root impact and that after execution prices should revert to 2/3 of their peak value.   

The analogous quantity to metaorder size studied here is $\pi = Q/V$.  From Figure~\ref{fig_stat} it is clear that this is not distributed according to a power law\footnote{
One complication is that we only study metaorders that are executed within the course of a single day, which truncates the distribution.  Nonetheless, based on Figure~\ref{fig_stat} it seems unlikely that removing this truncation would restore a power law.}.
As a result, it is not clear what this model implies.  Further work is needed to fit a functional form to the distribution of $\pi$ and work out the predictions for market impact under the fair pricing principle, but this is beyond the scope of this paper.

\subsection{Other theories}

Several other theories deserve mention.  The theory of Gabaix et al. \cite{gabaix06}  also predicts a square root for market impact.  However, this theory requires a very strong assumption, namely that the utility function of investors has absolute risk aversion, i.e. they assume that investors have a utility function of the form $\mu - \sigma^\delta$, where $\mu$ is the mean of returns and $\sigma$ is the standard deviation and $\delta = 1$.  If $\delta = 2$, for example, then the impact becomes linear.  

In view of our results a theory that is particularly worthy of mention is the PhD thesis of Austin Gerig\footnote{
This was joint work with J.D. Farmer and F. Lillo.} \cite{gerigthesis}. 
This model was an historical precursor to the theory of Farmer et al. discussed above.  As they did, Gerig assumed the prices form a martingale and that the starting and stopping times of metaorders are observable, but made a different auxiliary assumption.  This theory deserves special mention because it is the only theory that we are aware of that predicts a logarithmic dependence for market impact.  

\subsection{A few caveats}

Our data has limitations and we should issue some caveats.  In our data it is not possible to observe the strategic intentions of the agents originating the metaorders.   There may be preferential biases that are invisible to us.  In particular, suppose that execution of buy metaorders is sometimes cancelled before completion if the price rises too much (or if selling if the price falls too much).  This will systematically bias the sample to make impact appear more concave.  Even if the true impact were a square root, this could make the measured impact more concave.  Nonetheless, such effects would have to be substantial, and it seems a bit surprising that they would result in such good agreement with a logarithmic functional form.

Another important caveat that should be mentioned is the normalization by daily volume.  We make the implicit assumption, which has been almost universally made in prior work, that liquidity is proportional to daily volume.  This provides a (time varying) point of reference for market impact.  This is an assumption, and is not part of the predictions of any of the fundamental theories discussed above. A failure of the core assumption that daily volume is the correct way to measure liquidity could easily distort the shape of the impact function.  The only exception to the above is Kyle's original 1985 model and the new Kyle-Obizhaeva market invariance model, which predict a more complicated liquidity scaling \cite{Kyle14}.  We have not tested any such alternatives.  

Finally we should remind the reader that we truncate all metaorders that are longer than one day in duration (so that a metaorder that persists for $n$ days is treated as $n$ separate metaorders).  However, our inspection of the data suggests that this is rare -- see Figure~\ref{fig_aapl}.

\section{Conclusions}\label{sec:conclusions}

We have presented the most extensive empirical analysis of the market impact of the execution of large trades performed so far, at least in terms of the number  of metaorders and heterogeneity of their originators that have been analyzed.  The large dataset allows us to reduce the statistical uncertainty in the analysis and thereby make stronger inferences about the functional form of market impact.  We have also linked together the raw data on metaorders with minute-by-minute data on prices, so that we can study time dependent effects, such as the immediate impact as a metaorder is executed and the reversion after it is completed.  Our results extend but also contrast with what is commonly believed about market impact. Some of our main conclusions are as follows:
\begin{itemize}
\item Market impact conditional on the daily fraction $\pi$ (the ratio between the volume and the average daily volume) is remarkably well described by a logarithmic function over more than four orders of magnitude. In contrast, the square root impact law, which is widely used in academia and industry, approximates market impact only for a couple of orders of magnitude in $\pi$.  Thus the form of market impact is strongly concave, even more so than suggested by the square root law.
\item The {\it market impact surface} captures an inherently bivariate dependence of impact on participation rate and duration. As before, this bivariate dependence is much better represented by a logarithm than by a power law.   Furthermore, the good "collapse" seen by conditioning on $\pi$ alone is substantially due to a compensation effect between residuals.  That is, we show that impact depends on $F$ and $\eta$ separately; however, when one aggregates by conditioning only on $\pi$, the dependences tend to cancel each other.
\item During execution the price trajectory deviates from the temporary market impact and sometimes the price starts reverting well before the end of the execution.  This strongly suggests that market impact is decaying even as the metaorder is being executed.  We believe the lack of correspondence is due to front-loaded execution.  (This also reflects a limitation of our analysis; we do not have detailed timestamps for the execution of the metaorder, and so we are forced to assume uniform execution).
\item The propagator model is only a good description of metaorder impact for metaorders with high participation rate.  This is likely due, at least in part, to the overlap (herding) between metaorders of the same sign, which for moderate to low participation rate modifies the price dynamics considerably. 
\item  Prices clearly show a strong tendency toward reversion after metaorder execution ends.   This behavior strongly depends on both participation rate and duration.  For high participation rates orders of all duration relax in essentially the same way, consistent with a  propagator in which impact relaxes to zero as the square root of time.  In contrast for low participation rates orders of different duration behavior quite differently, with orders of longer duration relaxing more slowly than those of short duration. 
\end{itemize}

Our results present several modeling challenges, since none of the available models are fully capable of explaining our results.  Indeed, the only model that we are aware of that predicts logarithmic behavior is one due to the work of Farmer, Gerig and Lillo, which is reported in Austin Gerig's thesis \cite{gerigthesis}.  These results are particularly surprising when compared to the work of \cite{toth2011anomalous}.  In that work they show that square root behavior of impact is a necessary condition for diffusive behavior of prices, and that deviations from this should therefore result in arbitrage.  This raises the question of how the observed logarithmic impact can avoid this problem.  Please note however the caveats given in the previous section.  At the very least our work suggests the need for more large studies of market impact.  Unless there are biases in our results as discussed above, our work suggests that current fundamental theories of market impact have serious problems and that models of market impact require further development.

\section*{Acknowledgments}

We thank E. Bacry, J.-P. Bouchaud, J. Gatheral, P. Kyle, and H. Waelbroeck for useful and inspiring discussions. EZ, MT, and FL wish to thank Enrico Melchioni for the constant encouragement. The opinions expressed here are solely those of the authors and do not represent in any way those of their employers.

\bibliography{bibliography_mi}
\bibliographystyle{ieeetr}
 
\end{document}